\documentclass[12pt]{imsart}
\usepackage{amsthm,amsmath}
\usepackage[colorlinks,citecolor=blue,urlcolor=blue]{hyperref}
\usepackage{amsmath,amsfonts,amssymb,amsthm}
\usepackage{enumitem}
\RequirePackage{amsthm,amsmath,amsfonts,amssymb}
\RequirePackage[numbers]{natbib}
\RequirePackage[colorlinks,citecolor=blue,urlcolor=blue]{hyperref}
\RequirePackage{graphicx}
\usepackage{lineno,hyperref}
\startlocaldefs
\modulolinenumbers[5]
\usepackage[utf8]{inputenc}
\usepackage[english]{babel}
\usepackage[margin=1in]{geometry}
\usepackage{bm}
\usepackage{graphicx}
\usepackage{amssymb,amsmath,amsthm,mathrsfs}
\usepackage{epstopdf}
\usepackage{algorithm}
\usepackage{amsfonts,delarray}
\usepackage{algorithm,algcompatible,amsmath}
\usepackage{rotating,color}
\usepackage{subfigure}
\usepackage{multirow}
\usepackage{titlesec,textcase}
\usepackage{longtable}
\usepackage{bbm}
\usepackage{rotating}

\usepackage{caption}
\newtheorem{Theorem}{Theorem}[section]

\newtheorem{Definition}[Theorem]{Definition}
\theoremstyle{definition}
\newtheorem{Example}{Example}[section]

\RequirePackage[normalem]{ulem} 

\definecolor{rp}{RGB}{83,54,106}

\def\boxit#1{\vbox{\hrule\hbox{\vrule\kern6pt\vbox{\kern6pt#1\kern6pt}\kern6pt\vrule}\hrule}}

\endlocaldefs

\allowdisplaybreaks

\begin{document}
\begin{frontmatter}
\title{Empirical Likelihood Test for Common Invariant Subspace of Multilayer Networks based on Monte Carlo Approximation}

\runtitle{EL test for common invariant subspace}
\runauthor{ }

\begin{aug}

\author[A]{\fnms{Qianqian} \snm{Yao}\ead[label=e1] {qianqian.yao@ndsu.edu}}

\address[A]{
Department of Statistics, North Dakota State University, Fargo, ND, USA \\
\printead{e1}}

\end{aug}

\begin{abstract}
Multilayer (or multiple) networks are widely used to represent diverse patterns of relationships among objects in increasingly complex real-world systems. Identifying a common invariant subspace across network layers has become an active area of research, as such a subspace can filter out layer-specific noise, facilitate cross-network comparisons, reduce dimensionality, and extract shared structural features of scientific interest. One statistical approach to detecting a common subspace is hypothesis testing, which evaluates whether the observed networks share a common latent structure. In this paper, we propose an empirical likelihood (EL) based test for this purpose. The null hypothesis states that all network layers share the same invariant subspace, whereas under the alternative hypothesis at least two layers differ in their subspaces. We study the asymptotic behavior of the proposed test via Monte Carlo approximation and assess its finite-sample performance through extensive simulations. The simulation results demonstrate that the proposed method achieves satisfactory size and power, and its practical utility is further illustrated with a real-data application.

\end{abstract}

\begin{keyword}[class=MSC2020]
\kwd[]{60K35}
\kwd[; ]{05C80}
\kwd[; ]{62G10}
\end{keyword}

\begin{keyword}
\kwd{multilayer graphs}
\kwd{common invariant subspace}
\kwd{empirical likelihood test}
\kwd{Monte Carlo Approximation}
\kwd{Simulation}
\end{keyword}

\end{frontmatter}

\section{Introduction}\label{S:1}

Graphs (or networks) are widely used data structures and serve as a common language for modeling connected data in complex systems. Fundamentally, a graph consists of a collection of nodes representing objects and a set of edges representing the interactions or relationships between pairs of these objects. Graph data can be found in a broad spectrum of application domains. For example, graphs can be applied to model social networks, where nodes represent individuals or entities, and edges typically denote friendships, collaborations, interactions, or other social ties \cite{Li21}. Graph models are also employed to model molecules in quantum chemistry, catalyst discovery, drug discovery, etc. to predict the properties of molecules.  The atoms in molecules are modeled as nodes and the bond between two atoms is modeled as edge \cite{Gilmer17, SS23, DTM20, Yao22}. Beyond these, networks are utilized in numerous other domains, including event graphs, computer networks, disease pathways, food webs, particle networks, underground transportation systems, economic and financial networks, communication systems, and so on. While some of these examples are widely recognized and others are more specialized, together they demonstrate the extensive and diverse applications of graph models in real-world systems.

\begin{figure}
\begin{center}
\includegraphics[width=6cm,height=6cm]{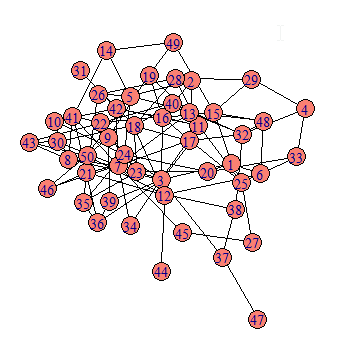} \hskip 0.1cm
\includegraphics[width=6cm,height=6cm]{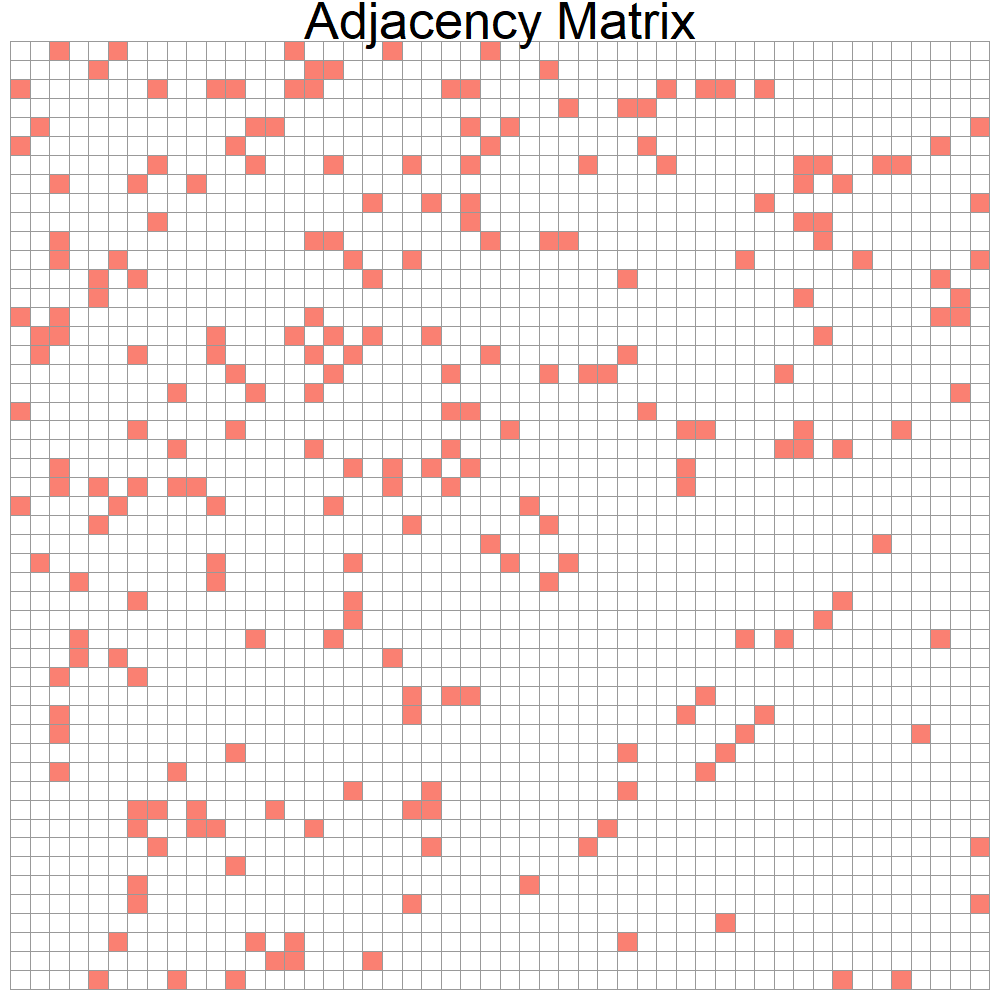}  
\end{center}
\caption{Network or Graph and its Adjacency Matrix: node size $n=50$}
\label{network1-n50}

\end{figure}

One of the most common mathematical representations of a graph is the adjacency matrix, where rows and columns correspond to the graph nodes, and the numerical values indicate the presence of edges between node pairs.
In this work, the nodes are indexed by \(1, 2, \dots, n\), and the adjacency matrix is of size \(n \times n\). We assume that the graph is unweighted, so the adjacency matrix contains only binary entries: 1 and 0. Specifically, $A_{ij} = 1$ if there is an edge between node $i$ and node $j$, and $A_{ij} = 0$ if there is no edge. All diagonal elements are set to 0, as self-connections (edges from a node to itself) are not considered. We also assume that the graph is undirected, meaning that we do not consider whether the edge is from node $i$ to node $j$ or from node $j$ to node $i$, as long as node $i$ and node $j$ are connected, $A_{ij} =1$ or $A_{ji} =1$. Thus, the adjacency matrix is symmetric. 
Figure~\ref{network1-n50} presents an example of a graph with 50 nodes and its corresponding adjacency matrix, which has dimensions of $50 \times 50$. The colored cells in the adjacency matrix represent the existence of edges. From the visualization, the adjacency matrix is symmetric and contains zeros on the diagonal.

Due to its widespread applications, graph data mining has gained tremendous popularity in the past decades. For instance, \cite{AV14,YS22} studied the sharp information-theoretic thresholds for testing the existence of dense subgraphs in  random graphs. \cite{BS16} designed a recursive bi-partitioning algorithm to detect community structure in networks. \cite{JKL21} proposed the optimal polygon test for testing community structure in heterogeneous networks. Most existing methods have been developed for single-layer networks. However, extending these approaches to multilayer networks---an increasingly important framework for modeling complex systems---poses new challenges. To address this, we turn to the topic of multilayer networks.

\begin{figure}
\begin{center}
\includegraphics[width=7cm,height=7cm]{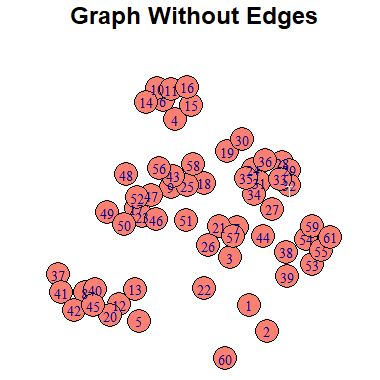} \hskip 10mm
\includegraphics[width=7cm,height=7cm]{{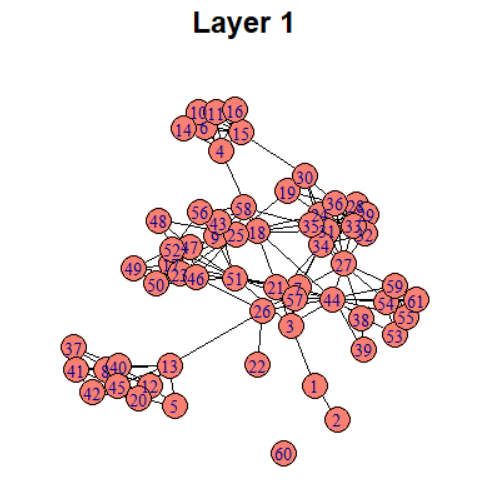}} 
\end{center}

\begin{center}
\includegraphics[width=7cm,height=7cm]{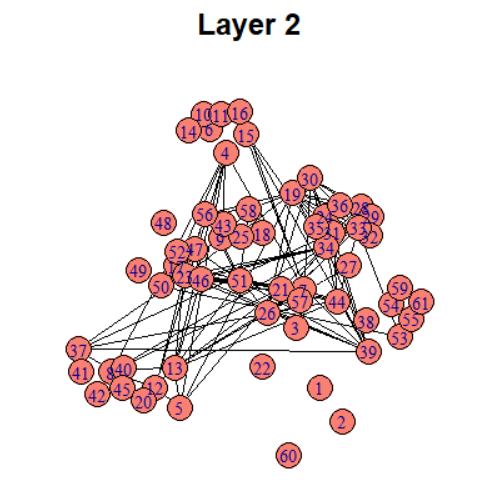} \hskip 10mm
\includegraphics[width=7cm,height=7cm]{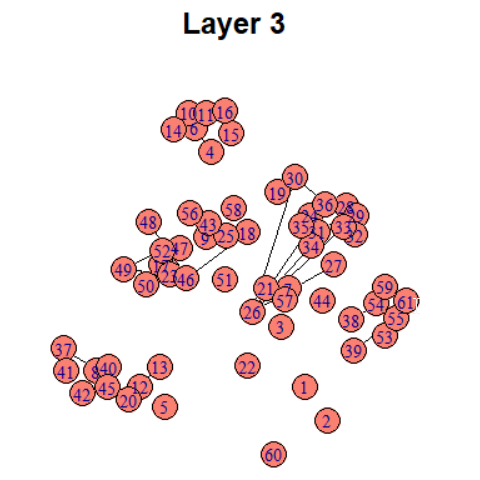} 
\end{center}

\begin{center}
\includegraphics[width=7cm,height=7cm]{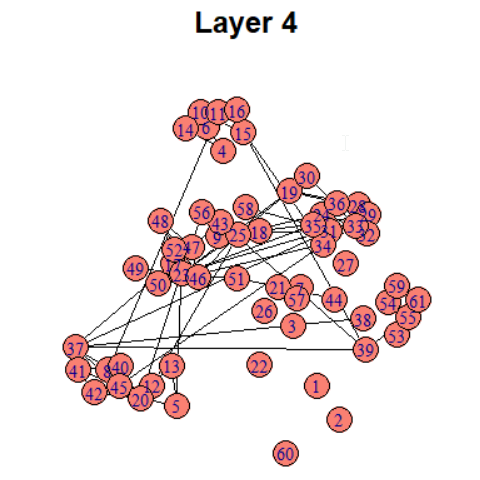} \hskip 10mm
\includegraphics[width=7cm,height=7cm]{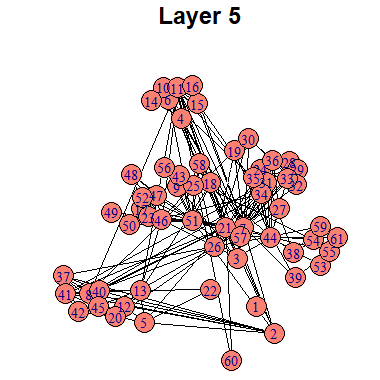} 
\end{center}

\caption{Five Layers Real-World CS-Aarhus Networks}
\label{cs-realdata1}
\end{figure}

A multilayer networks is a collection of networks that model complex systems by representing a fixed set of objects or entities as nodes, and capturing various types of relationships among them across different layers. Given a fixed set of nodes, each type of relationship is represented by a separate network layer. Together, these layers form the multilayer network, where each layer encodes a specific mode of interaction or connection among the same set of nodes. 
Multilayer networks are powerful tools for modeling multiple types of interactions that cannot be adequately captured by a single network or graph. For example, in multilayer social networks, one layer may represent personal friendships, another professional collaborations, and a third shared interests or activities \cite{OKH14}. A concrete example is the CS-Aarhus dataset described in \cite{MMR13}, which contains multilayer social networks of 61 employees in the computer science department at Aarhus University. These networks are constructed under assumptions of undirected, unweighted and no self-loops,  and include five types of online and offline relationships. Figure \ref{cs-realdata1} displays the five layers, each containing the same 61 nodes corresponding to the employees. Specifically, the first layer represents having lunch together, the second captures Facebook connections, the third represents co-authorship of publications, the fourth encodes leisure activities, and the fifth indicates working relationships.

There are several types of multilayer networks discussed in the literature. Edges between nodes within the same layer are referred to as intra-layer connections, while edges linking nodes across different layers are known as inter-layer connections. When inter-layer edges connect a node to its counterparts in other layers (i.e., nodes representing the same entity), the resulting structure is called a multiplex network. In contrast, if inter-layer edges connect nodes representing different entities across layers, the network is termed an interconnected network. In this work, we focus on multilayer networks (also referred to as multiple networks) that are defined on a common set of nodes, with edges occurring only within individual layers \cite{D17,DP15,LLK20,CLM22,Arroyo21, PW24, Pensky24, Kivela14, WTW22, SS16}. This setup corresponds to a multiplex network structure. This type of multiplayer networks find many applications in modeling real-world networks. For instance, in multilayer brain networks, the nodes represent brain regions and edges in each layer may encode activity in different frequency bands, activity of different tasks, and  functional connectivity \cite{D17}. Application of multilayer networks in biomedicine is discussed thoroughly in \cite{HK20}. An overview of multilayer network analysis and its application to epidemiological research questions has been proposed in \cite{KRSV20}. The approach of multilayer networks is also applied to study and quantify animal behavior through multifaceted networked systems \cite{FSPP19}.

Multilayer networks is a collection of networks used to model a community or system, where each layer represents a different mode of interaction within that system. Since each layer provides a distinct perspective on the same underlying set of entities, identifying common structural characteristics across all layers is a valuable approach for gaining deeper insights into the system or community as a whole.
In network analysis, community detection seeks to identify groups of nodes (communities) that are densely connected within themselves but sparsely connected to other groups. In the context of multilayer networks, a common invariant subspace refers to a set of nodes that exhibit consistent structural patterns across all layers. That is, nodes belonging to the same community in one layer are likely to belong to the same community in other layers. Therefore, identifying a common invariant subspace in multilayer networks can be interpreted as uncovering a shared community structure across different layers. Detecting such a common invariant subspace allows us to identify nodes that are stable and exhibit similar roles or characteristics throughout the multilayer structure.
Mathematically, the common invariant subspace of multilayer networks refers to a shared latent space that captures the underlying structural patterns and interactions consistently across all layers. This common invariant subspace acts as a unified representation, integrating the heterogeneous relational information embedded in each layer. This facilitates joint analysis and learning by integrating the multilayer structure into a coherent framework. Moreover, the common invariant subspace can serve as a basis for measuring similarity between multilayer networks. By comparing networks based on their shared subspace structure, one can cluster or classify multilayer networks into groups in which the constituent layers exhibit similar latent patterns.

\begin{figure}
\begin{center}
    \includegraphics[width=5cm,height=5cm]{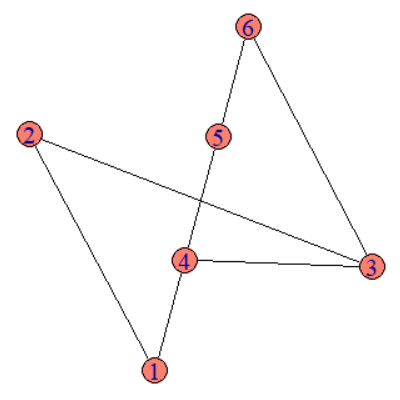} \hskip 5mm   \includegraphics[width=5cm,height=5cm]{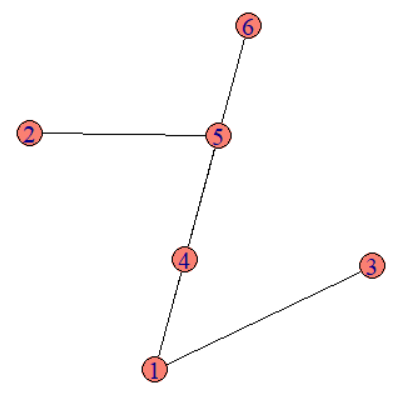} \hskip 5mm \includegraphics[width=5cm,height=5cm]{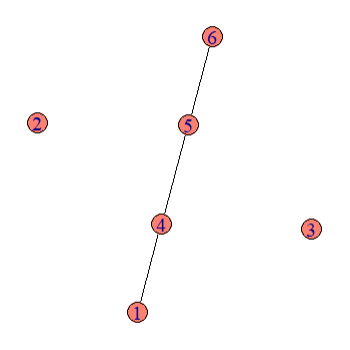} 
\end{center}
\caption{3 Layers Multilayer Networks: share common subspace}
\label{3Lcommonsubspace1}
\end{figure}

\begin{figure}
\begin{center}
    \includegraphics[width=5cm,height=5cm]{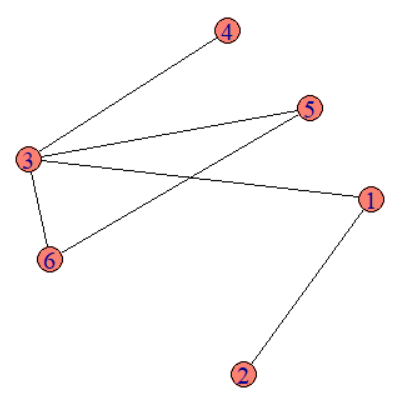} \hskip 5mm   \includegraphics[width=5cm,height=5cm]{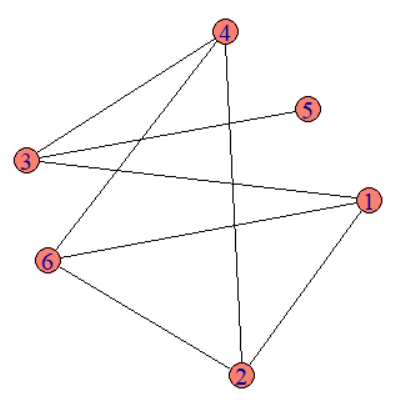} \hskip 5mm \includegraphics[width=5cm,height=5cm]{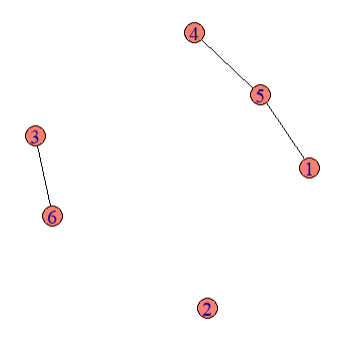} 
\end{center}
\caption{3 Layers Multilayer Networks: do not share common subspace}
\label{3Lcommonsubspace3}
\end{figure}

Figure \ref{3Lcommonsubspace1} illustrates an example of a three-layer multilayer network with six nodes, where nodes $1$, $4$, $5$, and $6$ are consistently connected across all layers. These nodes form a common subspace or common community shared by the three layers. In contrast, Figure \ref{3Lcommonsubspace3} demonstrate multilayer networks in which no common subspace exists, as there is no group of nodes that exhibit consistent structural relationships across all three layers.
However, as network size increases, identifying a common subspace in multilayer networks through visualization alone becomes increasingly challenging. Thus, to develop rigorous methodologies for analyzing the common subspace in multilayer networks is an active research area.

There are some work has been done on common subspace analysis in multilayer networks, such as \cite{AP04, ZT22, Wang18, Chen24, Arroyo21, PW24, Pensky24, yuan-yao2025testing}. Given a collection of networks, Arroyo et al. 2021 \cite{Arroyo21} assume that all the networks share common subspace; Pensky and Wang 2024 \cite{PW24, Pensky24} assume that some of the networks share common subspace and some do not, and the networks can be partitioned into clusters such that the networks within the same cluster have the common subspace. However, in reality, it is unknown whether all the networks share common invariant subsapce or some of the networks in multilayer networks share the common subspace. In detecting a common subspace in multilayer networks, hypothesis testing provides a principled statistical framework for assessing whether such a shared structure exists. This approach was developed in our earlier work \cite{yuan-yao2025testing}. The null hypothesis states that all network layers share the same common invariant subspace, whereas the alternative hypothesis posits that at least two layers do not share a common invariant subspace.

Empirical Likelihood (EL), first introduced by Owen~\cite{owen1988, owen2001}, is a nonparametric method of statistical inference that combines the flexibility of likelihood-based methods with the robustness of empirical data. Empirical Likelihood method is also discussed in \cite{lazar2021EL, liu2023EL}. Unlike traditional parametric likelihood approaches, empirical likelihood does not require specification of the underlying data distribution. Instead, it constructs a likelihood function directly from the observed data using probability weights subject to empirical constraints.
Due to its nonparametric nature, EL retains good efficiency while being robust to model misspecification, making it a powerful tool in modern statistical applications in various fields, including biostatistics, econometrics, and machine learning \cite{ding2019EL, newey2004EL, peng2013EL, ravuri2023EL, zhao2012EL}.

In this paper, we propose the Empirical Likelihood (EL) test based on the model proposed our earlier work \cite{yuan-yao2025testing}. We study the asymptotic behavior of the proposed test via Monte Carlo approximation and examine its finite-sample performance through extensive simulations. The Empirical Likelihood Test (EL test) demonstrates higher power than the Weighted Degree Difference Test (WDDT) in \cite{yuan-yao2025testing} under many certain same simulation settings, showing advantages. Its practical utility is further illustrated with a real-data application. By expanding the scope of empirical likelihood methods, this paper also strengthens the methodological toolkit available to statisticians and data scientists.

The paper is organized as follows. Section~\ref{mainRe} formally introduces the models and hypotheses. Section~\ref{ELtest} proposes the empirical likelihood (EL) test. Section~\ref{montecarlo} presents the Monte Carlo approximation of the limiting distribution. Section~\ref{SimRD} reports simulation results and examines how characteristics of multilayer networks affect the test performance and the robustness of the test. Section~\ref{realdata} illustrates the practical utility of the proposed method using a real-data application. Finally, Section~\ref{discuss} provides a discussion of the results and conclusions.

\vskip 5mm
\noindent
{\bf Notation:} We adopt the  Bachmann–Landau notation throughout this paper. Let $a_n$  and $b_n$ be two positive sequences. Denote $a_n=\Theta(b_n)$ if $c_1b_n\leq a_n\leq c_2 b_n$ for some positive constants $c_1,c_2$. Denote  $a_n=\omega(b_n)$ if $\lim_{n\rightarrow\infty}\frac{a_n}{b_n}=\infty$. Denote $a_n=O(b_n)$ if $a_n\leq cb_n$ for some positive constants $c$. Denote $a_n=o(b_n)$ if $\lim_{n\rightarrow\infty}\frac{a_n}{b_n}=0$. Let $X_n,X$ be random variables. Denote $X_n=O_P(a_n)$ if $\frac{X_n}{a_n}$ is bounded in probability. Denote $X_n=o_P(a_n)$ if $\frac{X_n}{a_n}$ converges to zero in probability as $n$ goes to infinity. Let $\mathbb{E}[X]$ and $Var(X)$ denote the expectation and variance of a random variable $X$ respectively. 
For positive integer $n$,$i,j,k$, denote $[n]=\{1,2,\dots,n\}$, and $i\neq j\neq k$ means $i\neq j, j\neq k, k\neq i$. Given positive integer $t$, $\sum_{i_1\neq i_2\neq\dots\neq i_t}$ means summation over all integers $i_1,i_2,\dots,i_t$ in $[n]$ such that $|\{i_1,i_2,\dots,i_t\}|=t$. $\sum_{i_1< i_2<\dots< i_t}$ means summation over all integers $i_1,i_2,\dots,i_t$ in $[n]$ such that $i_1<i_2<\dots<i_t$. For a vector $W=(W_1,W_2,\dots,W_m)\in\mathbb{R}^m$ and a positive integer $q$, $||W||_q=\left(\sum_i^m|W_i|^q\right)^{\frac{1}{q}}$.

\section{Model and Hypothesis}\label{mainRe}

We study multilayer networks where all layers have the same set of nodes and edges only connect nodes within each layer \cite{D17,DP15,LLK20,CLM22,Arroyo21, PW24, Pensky24}. Specifically, the multilayer networks consists of  $L$ graphs $G_1,G_2,\dots,G_{L}$, with $G_{l}=(\mathcal{V},\mathcal{E}_{l})$, where $\mathcal{V}=\{1,2,\dots,n\}$ is the node set, and $\mathcal{E}_{l}$ denotes a set of edges in graph $G_{l}$ \footnote{$G_{l}$ or $G^{(l)}$, $A_{l}$ or $A^{(l)}$ are the notation used to indicate the $l$-th layer graph or adjacency matrix in multilayer networks in this work.}. 
Assume the graphs in each layer are undirected and unweighted, without self-loops. 
Each graph \(G_l\) is represented by an \(n \times n\) symmetric adjacency matrix \(A_l\), where \(A_{l,ij} = 1\) if nodes \(i\) and \(j\) are connected by an edge in layer \(l\), and \(A_{l,ij} = 0\) otherwise. Additionally, all diagonal elements satisfy \(A_{l,ii} = 0\), indicating no self-loops.
A typical example of the multilayer networks is brain networks, where nodes represent brain regions, and edges model interactions between two brain regions \cite{D17,DP15,SBL13}. Multilayer brain networks have the same nodes, and there is no edge connecting nodes of different networks.

This section presents the related models and proposes a model for testing the common invariant subspace. We first consider a random heterogeneous graph model for a single network, as defined in Definition~\ref{random1}.

\begin{Definition}[Random Heterogeneous Graph 
(Erd\H{o}s--R\'enyi 1960; Bollob\'as et al. 2007)]\label{random1}

A \emph{random heterogeneous graph} is defined by the adjacency matrix $A = (A_{ij})$, where
\[
A_{ij} \stackrel{\text{i.i.d.}}{\sim} \text{Bernoulli}(P_{ij}), \quad \text{for } i < j,
\]
with $P_{ij} \in [0, 1]$. Here, $P = (P_{ij})$ is an $n \times n$ probability matrix, and set $P_{ii} = 0$.
The adjacency matrix $A$ satisfies the symmetry properties:
\[
A_{ij} = A_{ji}, \quad A_{ii} = 0,
\]
and its upper-triangular entries $A_{ij}$ (for $i < j$) are mutually independent.

\end{Definition}

By assumption of undirected and unweighted graph, the adjacency matrix satisfies $A_{ij} = A_{ji}$. Additionally, assumption of no self-loops implies $A_{ii} = 0$ for all $i$. Some random graph models may generate a probability matrix $P$ which contains nonzero diagonal entries. To ensure consistency with the graph assumption that $A_{ii} = 0$, we set the constraint that the expected value of each diagonal entry satisfies $\mathbb{E}[A_{ii}] = P_{ii} = 0$ in Definition \ref{random1}. Therefore, throughout this work, we impose the following model assumption on random graph model:
\[
P_{ii} = 0, \quad \text{for all } i \in [n].
\]

The \emph{Bernoulli random graph} model was first proposed by Erd\H{o}s and R\'enyi in \cite{erdos1959random, erdos1960evolution}, and is therefore also known as the \emph{Erd\H{o}s--R\'enyi random graph} model. In this model, edge probabilities are defined through a probability matrix $P$ of the same dimension as the adjacency matrix $A$.
A generalization of the \emph{Erd\H{o}s--R\'enyi random graph} was introduced in \cite{bollobas2007phase} and further discussed in \cite{lovasz2012large}, in which the probability of an edge between nodes $i$ and $j$ is not constant (i.e., heterogeneous), but instead specified by a matrix $P = (P_{ij})$. This class of models is commonly referred to as \emph{random heterogeneous graph} model.
Each layer in a multilayer networks can be modeled using the \emph{random heterogeneous graph} model. Accordingly, the random graph model for multilayer networks is constructed by applying the \emph{random heterogeneous graph} independently to each layer.

Before diving into random graph models for multilayer networks, some related terminologies in linear algebra are briefly discussed. For more detailed definitions, see \cite{rencher1997matrix, axler2024linear}. A \emph{subspace} is a subset $V$ of vector space in $\mathbb{R}^n$ and itself is also a finite-dimensional vector space. If the \emph{basis} of this finite-dimensional vector space is the list of vectors in $span\{ v_1, v_2, \dots, v_d\}$ and these vectors are \emph{linearly independent}, then the \emph{dimension} of this subspace is the number of these \emph{linearly independent} vectors. A list of vectors in $V$ is called \emph{linearly independent} if the only choice of $a_1, \dots, a_m \in \mathbb{R}$ that makes $a_1v_1 + \dots a_m v_m$ equal $0$ is $a_1 = \dots = a_m = 0$.

A linear map (also called linear transformation) from vector space $V$ to vector space $W$, denoted as $L(V, W)$, is a function $T: V \rightarrow W$ with the following properties: 1) additivity: $T(u+v) = T(u) + T(v)$ for all $u, v \in V$; 2) homogeneity: $T(\lambda v) = \lambda T(v)$. A \emph{Linear map} from a vector space to itself is denoted as $\mathcal{L}(V)$. For example, the \emph{identity map}, denoted as $I$, is the function on some vector space that takes each element to itself. To be specific, $I 
\in L(V)$ is defined by $Iv = v$.   
 Suppose a matrix $A \in \mathcal{L}(V)$, a \emph{subspace} $V$ is called \emph{invariant} under $A$ if $v \in V$ implies $Av \in V$. For example,  suppose $\lambda_1, \lambda_2, \dots, \lambda_d$ are distinct eigenvalues of $A \in \mathbb{R}^{n \times n}$ and $v_1, v_2, \dots, v_d$ are corresponding orthonormal eigenvectors, where $d \leq n$, then $v_1, v_2, \dots, v_d$ are \emph{linearly independent} and $V = span\{ v_1, v_2, \dots, v_d\}$ form a d-dimensional \emph{subspace} of $A$. By \emph{spectral theorem}, $AV = V\Lambda$, where $\Lambda = \text{diag}(\lambda_1, \dots, \lambda_d)$. Then, the subspace $V = span\{v_1, v_2, \dots, v_d\}$ is \emph{invariant} under $A$.

Suppose matrix $A$ is n-by-n matrix with entries in $\mathbb{R}$, the \emph{row rank} of $A$ is the dimension of the span of the \emph{linearly independent} rows, which is equivalently the number of the \emph{linearly independent} rows of $A$ in $\mathbb{R}^n$; the \emph{column rank} of $A$ is the dimension of the span of the \emph{linearly independent} columns, which is equivalently the number of the \emph{linearly independent} columns of $A$ in $\mathbb{R}^n$. The rank of a matrix $A \in \mathbb{R}^{n \times n}$ is the column rank of $A$.

Given an $L$-layer multilayer networks $A_1, A_2, \dots, A_L$ in which each layer contains $n$ nodes, each layer is represented by a symmetric adjacency matrix $A_l \in \mathbb{R}^{n \times n}$ for $l = 1, \dots, L$ that obtained based on \emph{random heterogeneous graph} model. Eigen-decomposition for each $\mathbb{E}[A_{l}]$:
\[
\mathbb{E}[A_{l}] V_l = V_l \Lambda_l, 
\quad \text{or equivalently,} \quad
\mathbb{E}[A_{l}] = V_l \Lambda_l V_l^\top,
\]
where $V_l \in \mathbb{R}^{n \times d}$ is an orthogonal matrix of eigenvectors representing a d-dimensional subspace of $\mathbb{E}[A_{l}]$,
and $\Lambda_l \in \mathbb{R}^{d \times d}$ is a diagonal matrix of eigenvalues, where $d \leq n$. For multilayer networks, this decomposition need to be further generalized to capture the underlying structure that is consistent among all layers. 
The following models are specifically tailored for multilayer networks.

\begin{Definition}[COmmon Subspace Independent Edge graphs (Arroyo et al. 2021)] \label{Arroyorandomgraph}

Let $V = (V_1, V_2, \ldots, V_n)^\top \in \mathbb{R}^{n \times d}$ be a matrix with orthonormal columns, and let $R_1, \ldots, R_{L} \in \mathbb{R}^{d \times d}$ be symmetric matrices such that
\[
0 \leq V_i^\top R_{l} V_j \leq 1 \quad \text{for all } i, j \in [n], \; l \in [L].
\]
Then the random adjacency matrices $A_1, \ldots, A_L$ are jointly distributed according to the \emph{COmmon Subspace Independent Edge} (COSIE) graph model with rank $d$ and parameters $V$ and $R_1, \ldots, R_L$ if, for each layer $l = 1, \ldots, L$, and given $V$ and $R_l$, the entries of $A_l$ are independent and follow
\[
\mathbb{P}(A_l) = \prod_{i < j} \left( V_i^\top R_{l} V_j \right)^{A^{(l)}_{ij}} 
\left( 1 - V_i^\top R_{l} V_j \right)^{1 - A^{(l)}_{ij}}.
\]
Equivalently, letting $P_{ij}^{(l)} = V_i^\top R_l V_j$, the model can be expressed as
\[
\mathbb{P}(A_l) = \prod_{i < j} \left( P_{ij}^{(l)} \right)^{A^{(l)}_{ij}} 
\left( 1 - P_{ij}^{(l)} \right)^{1 - A^{(l)}_{ij}}.
\]
The joint distribution of the multilayer adjacency matrices is written as
\[
A_1, \ldots, A_L \sim \text{COSIE}(V; R_1, \ldots, R_L),
\]
where $V \in \mathbb{R}^{n \times d}$ is the shared latent subspace and each $R_{l} \in \mathbb{R}^{d \times d}$ is a layer-specific score matrix.

\end{Definition}

The \emph{common subspace independent edge graphs model (COSIE)} was developed in \cite{Arroyo21}. In this work, the authors proposed a model for multiple heterogeneous networks, where each network shares a common latent subspace structure, and they introduced a spectral algorithm to estimate the common invariant subspace across multilayer networks. Consider a collection of $L$ networks with adjacency matrices $A_{l} \in \{0, 1\}^{n \times n}$, for $1 \leq l \leq L$, where each edge is modeled as an independent Bernoulli random variable, the expected adjacency matrix for the $l$-th network can be decomposed as
\[
\mathbb{E}[A_{l}] = V R_{l} V^\top,
\]
where $V \in \mathbb{R}^{n \times d}$ is a matrix with orthonormal columns representing a d-dimensional subspace, shared across all networks. The matrix $R_{l} \in \mathbb{R}^{d \times d}$ is a graph-specific score matrix that may vary with each network and is not necessarily diagonal. This formulation allows each network to have its own expected edge structure while capturing shared latent subspace information.

\begin{Definition}[DIverse MultiPLEx Generalized Dot Product Graph (Pensky and Wang 2024)]\label{Penskyrandomgraph}

Consider an $L$-layer networks on the same set of $n$ vertices $[n] = \{ 1, ..., n \}$, where the tensor of probabilities of connection $\mathcal{P} \in [0,1]^{n \times n \times L}$ is formed by layers $P^{(l)}, l \in [L]$, that can be partitioned into M groups with the common subspace structure or community assignment. 
Given a label function $c: [L] \rightarrow [M]$, The probability matrices $P^{(l)}$ for $l \in [L]$ are given by
\[
P^{(l)} = V^{(m)} Q^{(l)} \left(V^{(m)}\right)^\top, \quad \text{where } m = c(l), \; m \in [M],
\]
where $Q^{(l)} = \left(Q^{(l)}\right)^\top$ and $V^{(m)}$ are matrices with orthonormal columns, such that all entries of $P^{(l)}$ lie in the interval $[0,1]$.

\end{Definition}

Definition~\ref{Penskyrandomgraph} introduces the \emph{DIverse MultiPLEx Generalized Dot Product Graph (DIMPLE-GDPG)} model. The DIMPLE-GDPG model proposed in~\cite{PW24} generalizes the COSIE random graph model introduced in~\cite{Arroyo21}. The DIMPLE-GDPG model assumes that there are \(M\) distinct subspaces in the multilayer network, whereas the COSIE model assumes that all layers share a single common invariant subspace. 
The expected adjacency matrices in multilayer networks are decomposed as
\[
\mathbb{E}[A^{(l)}] = V^{(m)} Q^{(l)} \left(V^{(m)}\right)^\top,
\]
where \( V^{(m)} \in \mathbb{R}^{n \times d} \) is the homogeneity matrix representing the m-th d-dimensional subspace and \( Q^{(l)} \) is the heterogeneity score matrix for layer\( \ l\). Each adjacency matrix \( A^{(l)} \) has a distinct heterogeneity component \( Q^{(l)} \), indicating that the multilayer networks is heterogeneous.

In this work, we aim to test whether there is a shared common invariant subspace or there are distinct subspaces in multilayer networks. We begin with the simplest case by assuming that $V$ is a vector to represent a 1-dimensional subspace. Accordingly, we propose the \emph{rank-1 degree-corrected random graph} model for multilayer networks in \cite{yuan-yao2025testing}, based on Definition~\ref{Arroyorandomgraph} and Definition~\ref{Penskyrandomgraph}, under the assumption that the dimension of homogeneity matrix $V$ is one. The detailed formulation is provided in Definition~\ref{defmg}.

\begin{Definition}[Rank-1 Random Multilayer 
 Heterogeneous Graphs (Yuan and Yao 2025)]\label{defmg}
 
 Given a positive integer $L$,
let $W_{l}$ be a vector in $[0,1]^n$ such that $\|W_{l}\|_2=1$ for all $l\in[L]$, and $\rho_{l}$ be a positive sequence that may depend on $n$. We say the multilayer networks $A_1,A_2,\dots, A_{L}$ follow the Random Multilayer Heterogeneous Graphs Model $\mathcal{G}_{n}(W_1,W_2,\dots,W_L)$ if
\begin{equation}\label{theproposedmodel} 
 \mathbb{P}(A_{l,ij}=1)=\rho_{l}W_{l;i}W_{l;j},\ \ \ i<j,
\end{equation}
where $A_{l,ii}=0$, $A_{l,ij}=A_{l,ji}$, $A_{l,ij}$ ($1\leq i<j\leq n$, $1\leq l\leq L$) are independent.

\end{Definition}

Definition \ref{defmg} defines multilayer networks in which each layer is the rank-1 degree-corrected Erd\H{o}s--R\'{e}nyi random graph. The expected degree of node $i$ in $A_{l}$ is proportional to $W_{l;i}$. Hence, $W_{l}$ is a vector of the \emph{degree-correction parameters} of $A_{l}$. 
\emph{Rank} of matrix A is the \emph{column rank} which is the dimension of the \emph{span} of the \emph{linearly independent} columns or equivalently the number of \emph{linearly independent} columns of $A$ in $\mathbb{R}^n$ \cite{rencher1997matrix, axler2024linear}. 
The adjacency matrix A is the \emph{outer product} of two vectors:
\begin{equation}\label{explainforrank1}
\mathbb{E} [A_{l}] = \rho_{l} W_{l} W_{l}^\top.
\end{equation}
which can be equivalently written as: \[\mathbb{E} [A_{l}] = [ \rho_l w_1 W_l,  \rho_l w_2 W_l, \dots,  \rho_l w_n W_l ], \] where $w_i$ are the elements in \textit{degree-correction parameter} $W_l$ vector \footnote{both $w_i$ and $W_{l,i}$ are used to indicate the elements in $W_l$.}.
Therefore, the adjacency matrix A is a \emph{rank-1 matrix}. Therefore, each network in all layers in $\mathcal{G}_{n}(W_1,W_2,\dots,W_{L})$ is a \emph{rank-1 degree-corrected Erd\H{o}s--R\'{e}nyi random heterogeneous graph}.


The single layer random heterogeneous graph $\mathcal{G}_{n}(W_{1})$ is also related to the popular Chung-Lu model in \cite{CL02}, where  $W_1$ is a vector of $n$ non-negative real numbers and $\rho_1=\frac{1}{\sum_{i=1}^nW_{1;i}}$. If we replace $W$ in \cite{CL02} by $\widetilde{W}=\frac{W}{\|W\|_2}$, then the Chung-Lu model \cite{CL02} is $\mathcal{G}_{n}(\widetilde{W})$. The rank-1 degree-corrected random graph has been widely used to model real-world networks \cite{CGL16,BFK18,YXL21}.

Our proposed model of random multilayer heterogeneous graphs $\mathcal{G}_{n}(W_1,W_2,\dots,W_{L})$ in Definition \ref{defmg} is related to the models defined in \cite{Arroyo21, PW24, Pensky24}. If $W_1=W_2=\dots=W_{L}=W$, then $\mathcal{G}_{n}(W_1,W_2,\dots,W_{L})$ is a special case of the common subspace independent edge random graphs model in \cite{Arroyo21}. In this case, the multilayer networks $A_{l}$ ($1\leq l\leq {L}$) share the same degree-correction parameters (or one-dimensional subspace) represented by $W$, and simultaneously  have sufficient heterogeneity due to distinct $\rho_{l}$.  
When some of the vectors $W_1,W_2,\dots, W_{L}$ are equal, $\mathcal{G}_{n}(W_1,W_2,\dots,W_{L})$ is a special case of the diverse multilayer networks model in \cite{PW24, Pensky24}. In this scenario, the multilayer networks $A_{l}$ ($1\leq l\leq L$) can be partitioned into clusters such that the networks within the same cluster have common degree-correction parameters (or one-dimensional subspace). \cite{PW24, Pensky24} presented several algorithms to recover the latent cluster and estimate the common subspaces in a general setting.

The estimation methods proposed in \cite{Arroyo21, PW24, Pensky24} rely on the assumption that some or all layers of the multilayer network share a common invariant subspace. However, in practice, it is generally unknown whether this assumption holds. In this work, we are the first to address this issue through a formal hypothesis testing framework. Specifically, we assume that the subspaces are one-dimensional and adopt the model described in Definition~\ref{defmg}.

Given multilayer networks $A_1,A_2,\dots, A_L\sim\mathcal{G}_{n}(W_1,W_2,\dots,W_L)$, we are interested in testing the following hypotheses
\begin{equation}\label{hypoeq}
H_0:W_1=W_2=\dots=W_L,\ \ \hskip 1cm H_1: W_{l_1}\neq W_{l_2},\ \text{for some }  l_1\neq l_2.
\end{equation}

Under $H_0$,  the graphs $A_1,A_2,\dots, A_L$ have  the same degree-correction parameters. Under $H_1$, there exist at least two graphs such that their degree-correction parameters are  different.

\section{Empirical Likelihood Test}\label{ELtest}

Empirical Likelihood (EL), first introduced by Owen~\cite{owen1988, owen2001}, is a nonparametric method of statistical inference that combines the flexibility of likelihood-based methods with the robustness of empirical data. Empirical Likelihood method is also discussed in \cite{lazar2021EL, liu2023EL}. Unlike traditional parametric likelihood approaches, empirical likelihood does not require specification of the underlying data distribution. Instead, it constructs a likelihood function directly from the observed data using probability weights subject to empirical constraints.
Due to its nonparametric nature, EL retains good efficiency while being robust to model misspecification, making it a powerful tool in modern statistical applications. In this chapter, we apply the EL framework to test common invariant subspace of multilayer networks, as formulated in~(\ref{hypoeq}), and examine its empirical performance through simulations.

\begin{Definition}[Owen 1988; Owen 2001; Lazar 2021]\label{defEL}

Given a random sample \( X_1, \dots, X_n \in \mathbb{R}^d \), where $d \geq 1$, from an unspecified distribution $F$ with mean $\mu \in \mathbb{R}^d$. Let $w_i$ be the weight that distribution function $F$ places on observation $X_i$. 
The ratio of the nonparametric likelihood $R(F)=\prod_{i=1}^{n} n w_i$, where $w_i \geq 0$ for all $i$ and $\sum_{i=1}^{n} w_i = 1$. Ties may or may not be present in sample. Then empirical likelihood ratio function for the mean is defined as
\[
R_n = \max \left\{ \prod_{i=1}^{n} n w_i \;\middle|\; \sum_{i=1}^{n} w_i X_i = \mu,\; w_i \geq 0,\; \sum_{i=1}^{n} w_i = 1 \right\}.
\]

\end{Definition}

Owen \cite{owen1988, owen2001} shows that the empirical likelihood ratio follows a chi-square distribution asymptotically, which is the usual asymptotic result for the parametric likelihood ratio test due to Wilks \cite{wilks1938large} and suggests that little is lost by the particular nonparametric shift represented by empirical likelihood.

\begin{Theorem}[Owen 1988; Owen 2001; Lazar 2021]\label{theoremEL} 

Let $X_1, X_2, ..., X_n \in \mathbb{R}^d$, where $d \geq 1$, be independent random variables with a common distribution $F_0$ with mean $\mu_0$ and finite variance covariance matrix $V_0$ of rank $q>0$. Then $-2 \text{log} R_n$ converges in distribution to a $\chi_{(q)}^2$ random variable as $n \rightarrow \infty$. 

\end{Theorem}


Based on Theorem \ref{theoremEL}, we define the Empirical Likelihood test (EL test) as follows:
\[\text{Reject $H_0$ at significance level $\alpha$, if $-2 \text{log}R_n>\chi^2_{q, 1-\alpha}$},\]
where \( \chi^2_{q, 1 - \alpha} \) denotes the upper \( \alpha \) critical value, or equivalently, the \( 100(1 - \alpha)\% \) percentile of the chi-squared distribution with q degree of freedom. Usually $q =d$, but if $q < d$, it means the $X_i$ are confined to a smaller subspace. Hence, the degrees of freedom in the limiting distribution adjust accordingly. Theorem~\ref{theoremEL} guarantees that the type I error of the EL test is asymptotically equal to \( \alpha \).

We propose the Empirical Likelihood Ratio Test (EL test) for the hypothesis testing problem (\ref{hypoeq}). Define the Weighted Degree Difference Data as 
\begin{equation}\label{dataEL}
X_i = \sum_{l=2}^{L}\left[\left(\frac{d_{1,i}}{\sqrt{P_1}}-\frac{d_{l,i}}{\sqrt{P_l}}\right)^2-\frac{d_{1}}{P_1}-\frac{d_{l}}{P_l}\right], 1\leq i \leq n. 
\end{equation}
where $P_{l}=\sum_{i\neq j\neq k}A_{l,ij}A_{l,jk}$, $d_{l;i}=\sum_{j}A_{l,ij}$ and $d_l=\sum_{i,j}A_{l,ij}$, for each $l\in[L]$.

In (\ref{dataEL}), the term $\frac{d_{1,i}}{\sqrt{P_1}}-\frac{d_{l,i}}{\sqrt{P_l}}$ is a weighted difference between degree $d_{1,i}$ of node $i$ in $A_1$ and $d_{l,i}$ of node $i$ in $A_l$. And $\frac{d_{l,i}}{\sqrt{P_l}}$ ($1\leq l\leq L$,$1\leq i\leq n$) are estimators of the parameters $W_{l;i}$ ($1\leq l\leq L$,$1\leq i\leq n$). 
Hence $\sum_{i=1}^n\left(\frac{d_{1,i}}{\sqrt{P_1}}-\frac{d_{l,i}}{\sqrt{P_l}}\right)^2$ measures the difference between $W_{1}$ and $W_{l}$.  The term $\frac{d_{1}}{P_1}+\frac{d_{l}}{P_l}$ centers  $\sum_{i=1}^n\left(\frac{d_{1,i}}{\sqrt{P_1}}-\frac{d_{l,i}}{\sqrt{P_l}}\right)^2$. 
Therefore, $X_i$ measure the sum of the node degree differences between $A_1$ and $A_l$ across the multilayer networks for $i \in n$. If the mean of $X_i$ equal to 0, which is under the null hypothesis, then the graphs $A_1,A_2,\dots, A_L$ have the same degree-correction parameters. Otherwise, there exist at least two graphs such that their degree-correction parameters are different.

The Weighted Degree Difference Data \( X_i \), defined in (\ref{dataEL}), represent the weighted degree differences for node \( i \), where \( 1 \leq i \leq n \), in multilayer networks. These data are not mutually independent, and hence it is unclear whether Theorem~\ref{theoremEL} still holds. The two examples below were also considered in \cite{yuan-yao2025testing}, our earlier work on testing the common invariant subspace of multilayer networks using the Weighted Degree Difference Test. The constraints imposed in these examples are designed to satisfy the theoretical assumptions required in \cite{yuan-yao2025testing} and are also expected to meet the conditions of Theorem \ref{theoremEL}. Prior to developing a rigorous mathematical proof of the limiting distribution, this study investigates the empirical distribution of the proposed empirical likelihood test statistic via Monte Carlo approximation, which is in Section~\ref{montecarlo}. The resulting empirical distributions provide numerical evidence supporting the validity of Theorem~\ref{theoremEL}. Simulation studies in Section \ref{SimRD} further investigate the Type~I error rates and the power of the proposed empirical likelihood test.

\begin{Example}\label{example0}
For positive constants $r,\lambda_l$ with $\lambda_l\leq1$ and $r>1$,
let $W_{l,i}=\frac{\lambda_l\sqrt{r}}{\sqrt{n}}$ for $1\leq i\leq \frac{n}{r}$ and $W_{l,i}=\frac{\sqrt{\frac{r}{r-1}(1-\lambda_l^2)}}{\sqrt{n}}$ for $\frac{n}{r}<i\leq n$. 
Then simple calculation yields 
\[\|W_l\|_1=\Theta(\sqrt{n}),\ \ \ \|W_l\|_2=1,\ \ \ \ \|W_l\|_4^4=O\left(\frac{1}{n}\right).\]        
If $\lambda_l=\lambda_1$, $\min_{1\leq l\leq L}\{\rho_l\}=\omega(1)$,  $\rho_l=o(\sqrt{n})$. 
Moreover, direct calculation yields
 \begin{eqnarray*}
\sum_{i=1}^nW_{1,i}W_{l,i}
&=&\lambda_1\lambda_l+\sqrt{(1-\lambda_1^2)(1-\lambda_l^2)}.
\end{eqnarray*} 
If $\lambda_l\neq\lambda_1$,  there exists a positive constant $\epsilon$ such that $\sum_{i=1}^nW_{1,i}W_{l,i}\leq 1-\epsilon$. The larger the difference between $\lambda_1$ and $\lambda_l$, the smaller the $\sum_{i=1}^nW_{1,i}W_{l,i}$.  
\end{Example}


\begin{Example}\label{example1}
    Let $m,\beta_l$ be non-negative constants for $l\in[L]$. Denote
    \[S_{n,m}=\sum_{i=1}^ni^{m}.\]
    When $m$ is a positive integer, $S_{n,m}$ is given by the Faulhaber's formula. For arbitrary positive constant $m$,
it is easy to verify that 
\begin{equation}\label{sumnm}
    S_{n,m}=\frac{n^{m+1}}{m+1}\left(1+o(1)\right).
\end{equation}

Let 
\[W_{l,i}=\frac{i^{\beta_l}}{\sqrt{S_{n,2\beta_l}}}.\]    
By the definition of $S_{n,m}$ and (\ref{sumnm}), one has
\[\|W_l\|_1=\sum_{i=1}^nW_{l,i}=\frac{1}{\sqrt{S_{n,2\beta_l}}}\sum_{i=1}^ni^{\beta_l}=\Theta\left(\sqrt{n}\right),\]
\[\|W_l\|_2^2=\sum_{i=1}^nW_{l,i}^2=\frac{1}{S_{n,2\beta_l}}\sum_{i=1}^ni^{2\beta_l}=1,\]        \[\|W_l\|_4^4=\sum_{i=1}^nW_{l,i}^4=\frac{1}{S_{n,2\beta_l}^2}\sum_{i=1}^ni^{4\beta_l}=O\left(\frac{1}{n}\right).\]
The expected degree of node $i$ in $A_l$ is $i^{\beta_l}\Theta\left(\frac{\rho_l}{n^{\beta_l}}\right)$. Hence the networks are highly heterogeneous.
If $\beta_l=\beta_1$, $\min_{1\leq l\leq L}\{\rho_l\}=\omega(1)$,   $\rho_l=o(\sqrt{n})$. 
In addition, direct calculation yields
\begin{eqnarray*}
\sum_{i=1}^nW_{1,i}W_{l,i}&=&\sum_{i=1}^n\frac{i^{\beta_1}}{\sqrt{S_{n,2\beta_1}}}\frac{i^{\beta_l}}{\sqrt{S_{n,2\beta_l}}}\\
&=&(1+o(1))\frac{n^{\beta_1+\beta_l+1}}{\beta_1+\beta_l+1}\frac{\sqrt{(2\beta_1+1)(2\beta_l+1)}}{\sqrt{n^{2\beta_1+2\beta_l+2}}}\\
&=&(1+o(1))\frac{\sqrt{(2\beta_1+1)(2\beta_l+1)}}{\beta_1+\beta_l+1}.
\end{eqnarray*}
If $\beta_l\neq \beta_1$, there exists a positive constant $\epsilon$ such that $\sum_{i=1}^nW_{1,i}W_{l,i}\leq 1-\epsilon$. The larger the difference between $\beta_1$ and $\beta_l$, the smaller the $\sum_{i=1}^nW_{1,i}W_{l,i}$.  

\end{Example}

\section{Monte Carlo Approximation of the Null Distribution}\label{montecarlo}

Multilayer networks are generated using Example~\ref{example0} and Example~\ref{example1}. In these Monte Carlo simulation settings, we know the ground truth about whether the generated multilayer networks shares a common subspace or not. 
We therefore investigate its finite-sample behavior through Monte Carlo simulation. Under the null hypothesis, multilayer networks are generated according to the prescribed model with a shared invariant subspace across layers. For each simulation, the test statistic is computed, and this procedure is repeated 10,000 times. The resulting empirical distribution provides an approximation to the null distribution of the test statistic, allowing us to evaluate its distributional properties and to obtain simulation-based critical values.

We first investigate the empirical distribution by Monte Carlo approximation based on the Example \ref{example0}. That is, the vectors $W_l=(W_{l,1},W_{l,2},\dots,W_{l,n})$ ($1\leq l\leq L$) are given by 
\[W_{l,i}=\frac{\lambda_l\sqrt{r}}{\sqrt{n}},\ \ \ 1\leq i\leq \frac{n}{r},\]
\[W_{l,i}=\frac{\sqrt{\frac{r}{r-1}(1-\lambda_l^2)}}{\sqrt{n}},\ \ \ \frac{n}{r}<i\leq n,\] 
where $0<\lambda_l\leq 1$ and $r>1$. The parameter \( \rho_l \) is defined as \( \rho_l = n^{\tau_l} \). Then network $A_l$ is generated by (\ref{theproposedmodel}), where $A_{l,ij}=A_{l,ji}$,  $A_{l,ij}$ ($1\leq i<j\leq n$, $1\leq l\leq L$) are independent. 

Set $r = 2$, $L =3$, $n = 400$. Denote $\boldsymbol{\tau}=(\tau_1,\dots,\tau_L)$ with \( \tau_l \in \{0.3, 0.2, 0.4, 0.1\} \) and $\boldsymbol{\lambda}=(\lambda_1,\dots,\lambda_L)$ with $\lambda_l\in\{0.8,0.7,0.6,0.5\}$. Each layer in the generated multilayer networks is characterized by two parameters:
\[A_l \sim \mathcal{G}_n(\tau_l, \lambda_l) \] and multilayer networks is indicated by its associated parameter vectors. 
For example, a three-layer networks can be denoted as:
\[(A_1, A_2, A_3) \sim \mathcal{G}_n(\boldsymbol{
\tau}, \boldsymbol{\lambda}) \quad \text{where } \boldsymbol{\tau} =(\tau_1, \tau_2, \tau_3), \boldsymbol{\lambda} =(\lambda_1, \lambda_2, \lambda_3). \]

\begin{figure}
\begin{center}
\includegraphics[width=8cm,height=6cm]{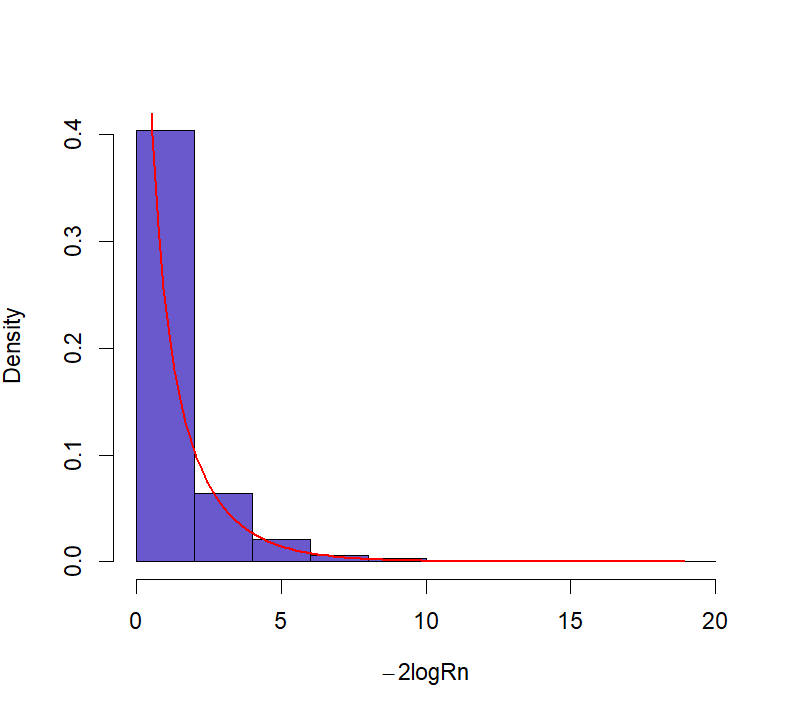} \hskip 0.1cm
\includegraphics[width=8cm,height=6cm]{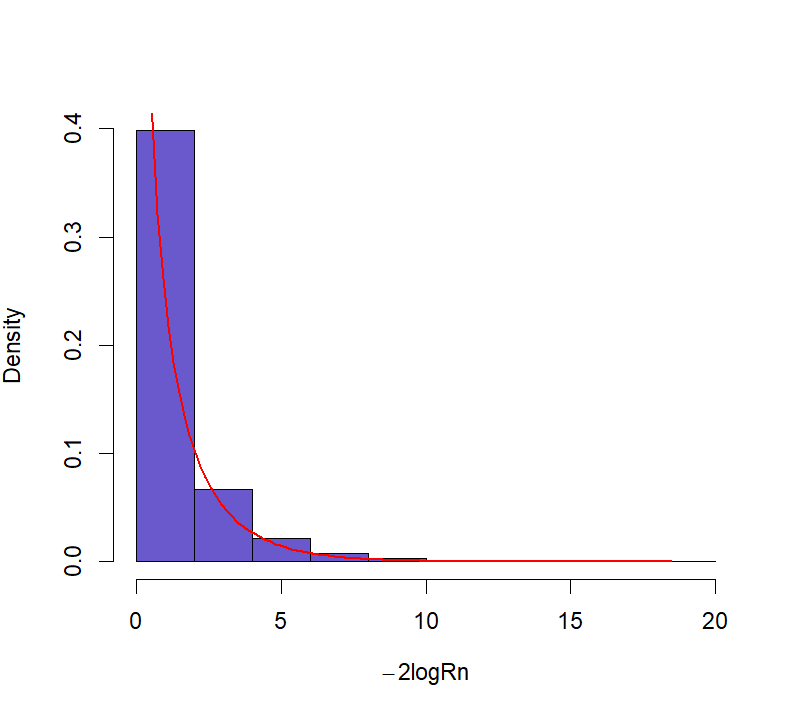}
\end{center}

$ \mathcal{G}_n(\boldsymbol{\tau}=(0.3, 0.2, 0.4), \boldsymbol{\lambda} =(0.8, 0.8, 0.8))$ \hskip 2cm $ \mathcal{G}_n(\boldsymbol{\tau}=(0.4, 0.3, 0.2), \boldsymbol{\lambda} =(0.8, 0.8, 0.8))$

\begin{center}
\includegraphics[width=8cm,height=6cm]{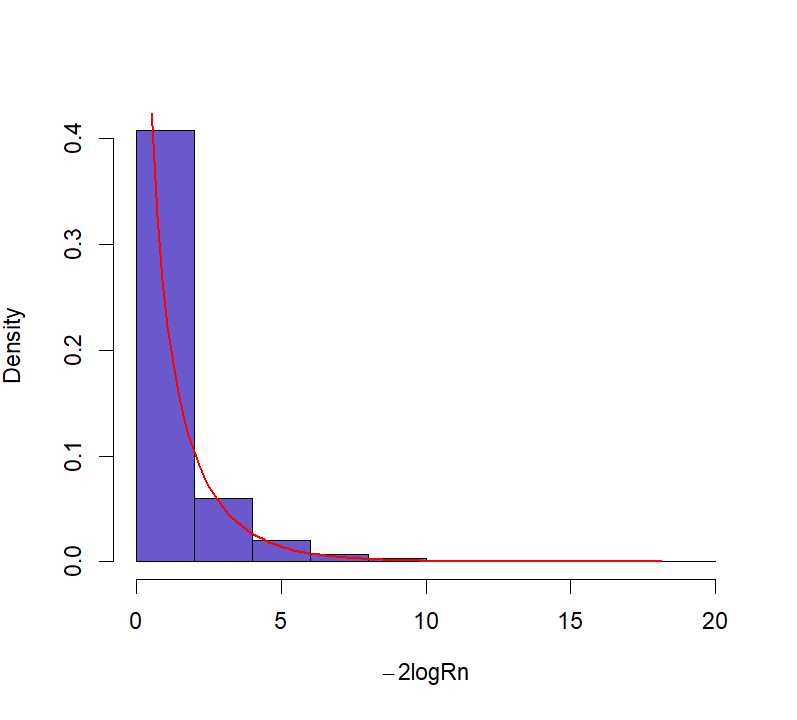} 
\end{center}

$ \mathcal{G}_n(\boldsymbol{\tau}=(0.2, 0.3, 0.4), \boldsymbol{\lambda} =(0.8, 0.8, 0.8))$

\caption[]{Monte Carlo Approximation Under Null Hypothesis}
\label{example1_null}
\end{figure}

\begin{figure}
\begin{center}
\includegraphics[width=8cm,height=6cm]{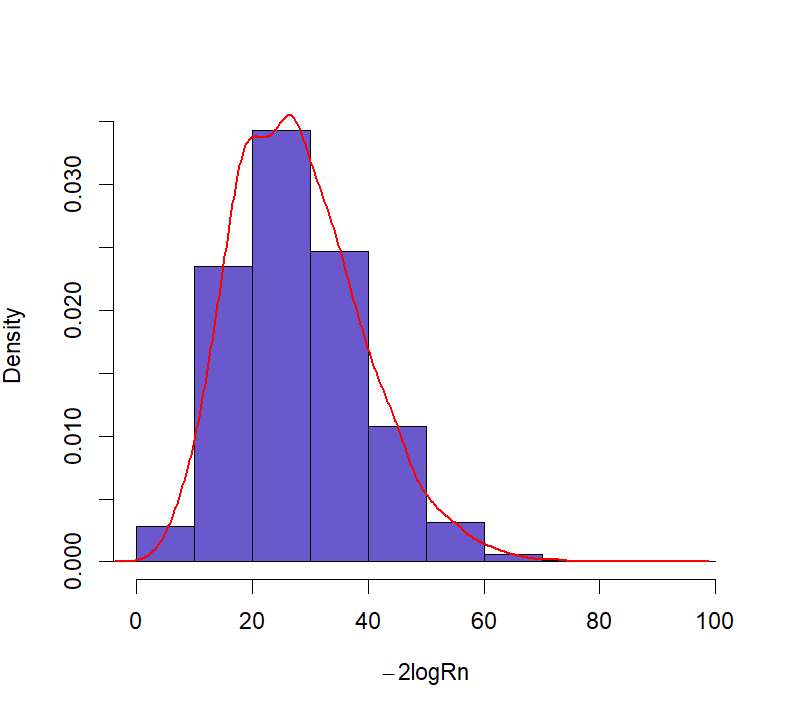} \hskip 0.1cm
\includegraphics[width=8cm,height=6cm]{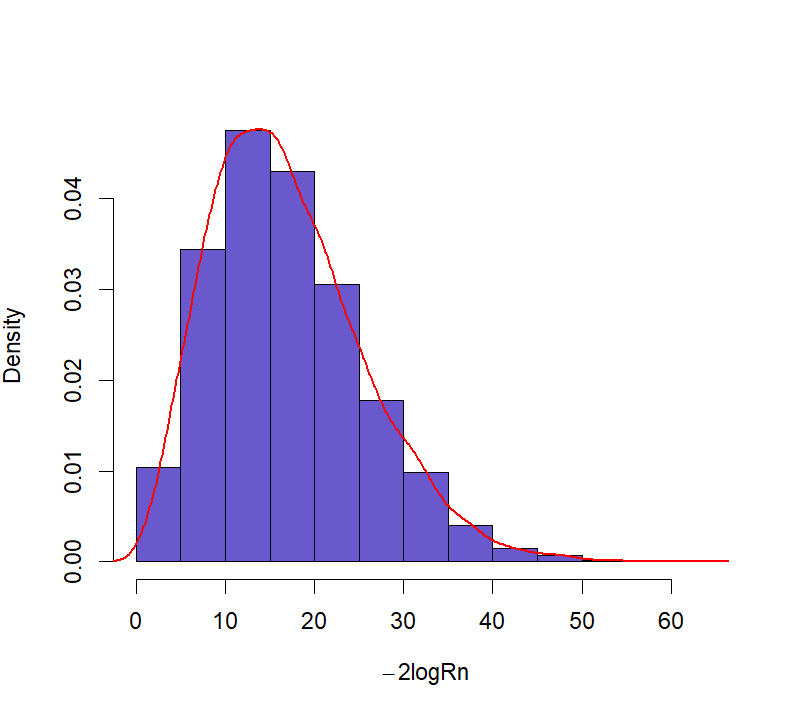}  
\end{center}

$\mathcal{G}_n(\boldsymbol{\tau}=(0.3, 0.2, 0.4), \boldsymbol{\lambda} =(0.8, 0.5, 0.5))$ \hskip 2cm $\mathcal{G}_n(\boldsymbol{\tau}=(0.3, 0.2, 0.4), \boldsymbol{\lambda} =(0.8, 0.6, 0.5))$

\begin{center}
\includegraphics[width=8cm,height=6cm]{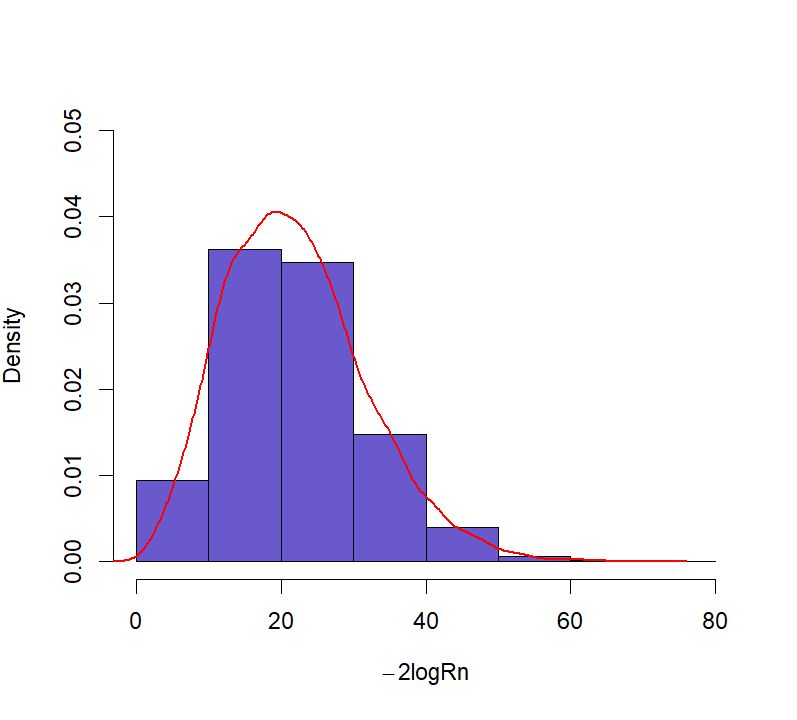} \hskip 0.1cm
\includegraphics[width=8cm,height=6cm]{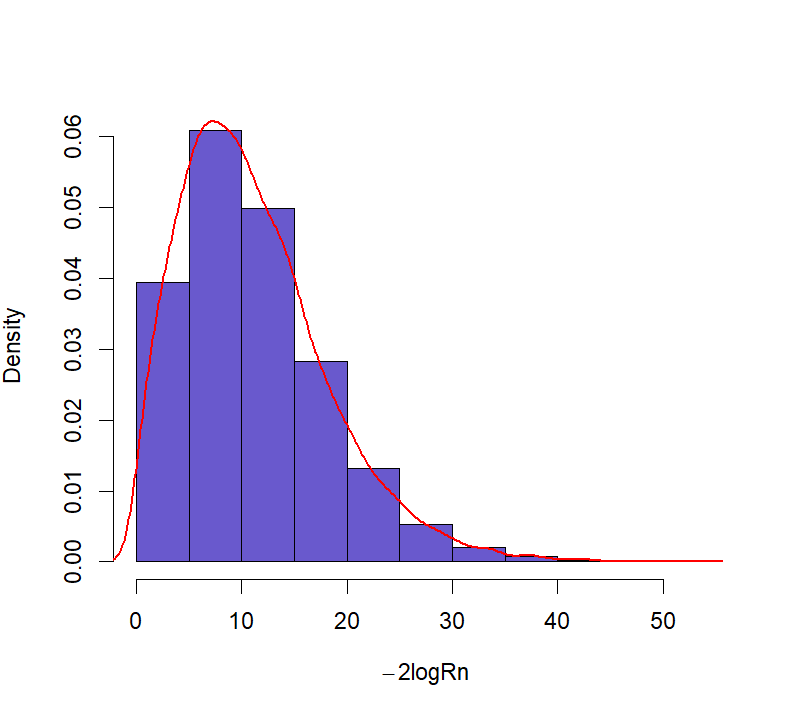}  
\end{center}

$ \mathcal{G}_n(\boldsymbol{\tau}=(0.4, 0.3, 0.2), \boldsymbol{\lambda} =(0.8, 0.6, 0.5))$ \hskip 2cm $\mathcal{G}_n(\boldsymbol{\tau}=(0.2, 0.3, 0.4), \boldsymbol{\lambda} =(0.8, 0.6, 0.5))$

\caption[]{Monte Carlo Approximation Under Alternative Hypothesis}
\label{example1_alt}
\end{figure}

To examine whether interactions between the parameters $\boldsymbol{\tau}$ and $\boldsymbol{\lambda}$ affect the testing performance, four scenarios are considered for three-layer networks, as summarized in Table~\ref{example1ThreeLayerPLOTscom} for Example~\ref{example0}. The first column presents three settings under the null hypothesis, while the second, third, and fourth columns correspond to settings under the alternative hypothesis. We first plot the empirical distributions for the three null-hypothesis settings in the first column and then examine the empirical distributions under the alternative-hypothesis setting in the fourth column.
The Monte Carlo approximations of the null distributions are presented in Figures~\ref{example1_null}. The red curves correspond to the chi-square distribution with one degree of freedom. The empirical density of the empirical likelihood test statistic closely aligns with the theoretical chi-square density, providing numerical evidence in support of Theorem~\ref{theoremEL}. Figure~\ref{example1_alt} presents the Monte Carlo approximations under the alternative hypothesis. Under the alternative, the EL test statistics exhibit an approximately normal distribution. This behavior is consistent with the Central Limit Theorem, since under fixed alternatives the test statistic converges to a normal distribution due to the accumulation of stochastic fluctuations from the estimating equations.


In the second simulation, the networks are generated from the model specified in Example \ref{example1}. That is, the vectors $W_l=(W_{l,1},W_{l,2},\dots,W_{l,n})$ ($1\leq l\leq L$) are given by 
\[W_{l,i}=\frac{i^{\beta_l}}{\sqrt{S_{n,2\beta_l}}},\] 
where $S_{n,m}=\sum_{i=1}^ni^{m}$, $m$ and $\beta_l$ are non-negative constants for $l\in[L]$. 

Let $\boldsymbol{\beta}=(\beta_1,\beta_2,\beta_3,\beta_4)$ with $\beta_l\in\{1,2,3,4\}$ and $\boldsymbol{\tau}=(\tau_1,\dots,\tau_L)$ with \( \tau_l \in \{0.3, 0.2, 0.4, 0.1\}. \) \( \rho_l \) is defined as \( \rho_l = n^{\tau_l} \).
Moreover, set $L =3$, $n=400$. 
Then generate $A_l$ according to (\ref{theproposedmodel}). 
Each layer in the generated multilayer networks is characterized by two parameters:
\[A_l \sim \mathcal{G}_n(\tau_l, \beta_l) \] and multilayer networks is indicated by its associated parameter vectors.  
For example, a three-layer networks can be denoted as:
\[(A_1, A_2, A_3) \sim \mathcal{G}_n(\boldsymbol{
\tau}, \boldsymbol{\beta}) \quad \text{where } \boldsymbol{\tau} =(\tau_1, \tau_2, \tau_3), \boldsymbol{\beta} =(\beta_1, \beta_2, \beta_3). \]

\begin{figure}
\begin{center}
\includegraphics[width=8cm,height=6cm]{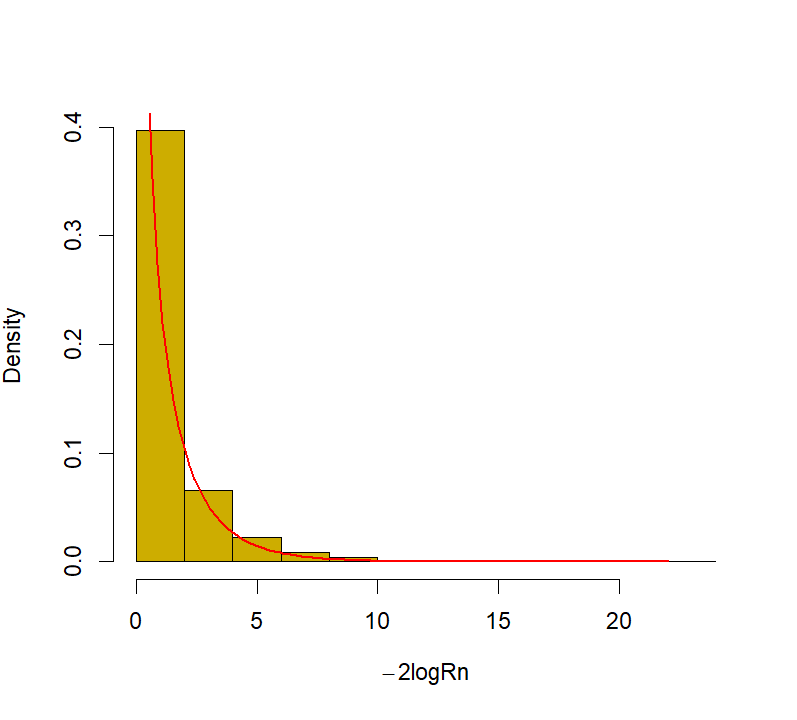} \hskip 0.1cm
\includegraphics[width=8cm,height=6cm]{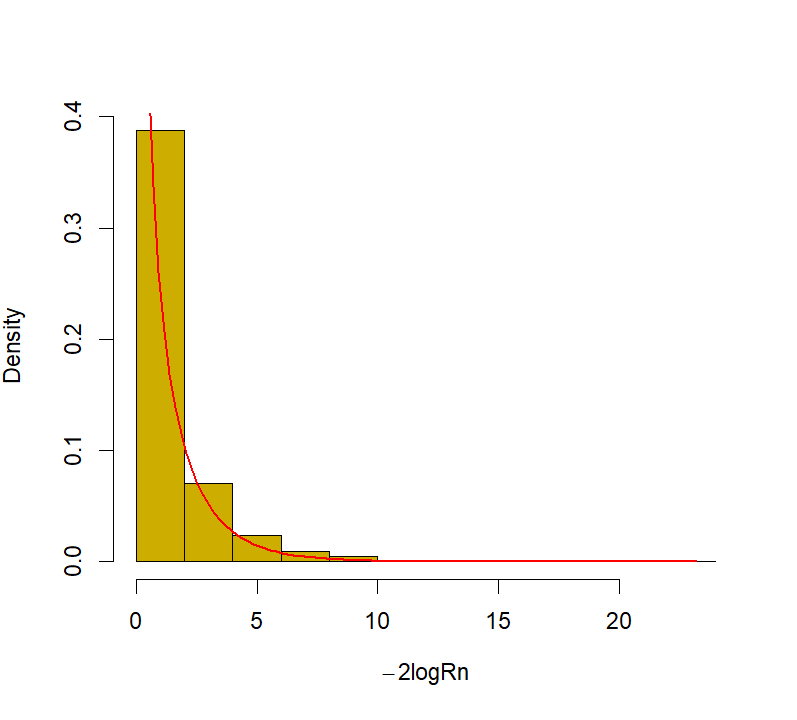}
\end{center}

$ \mathcal{G}_n(\boldsymbol{\tau}=(0.3, 0.2, 0.4), \boldsymbol{\lambda} =(1, 1, 1))$ \hskip 2cm $ \mathcal{G}_n(\boldsymbol{\tau}=(0.4, 0.3, 0.2), \boldsymbol{\lambda} =(1, 1, 1))$

\begin{center}
\includegraphics[width=8cm,height=6cm]{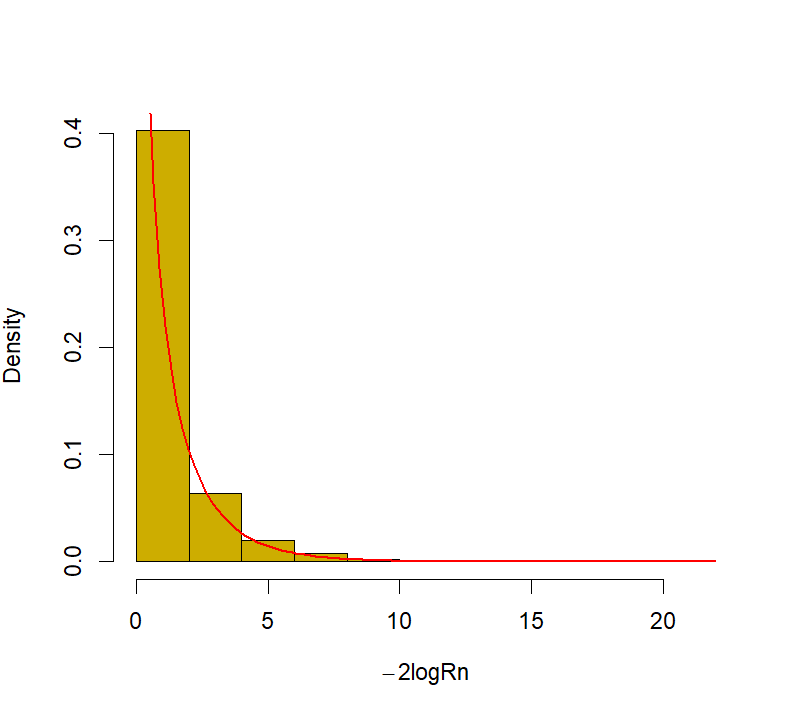} 
\end{center}

$ \mathcal{G}_n(\boldsymbol{\tau}=(0.2, 0.3, 0.4), \boldsymbol{\lambda} =(1, 1, 1))$

\caption[]{Monte Carlo Approximation Under Null Hypothesis}
\label{example2_null}
\end{figure}

\begin{figure}
\begin{center}
\includegraphics[width=8cm,height=6cm]{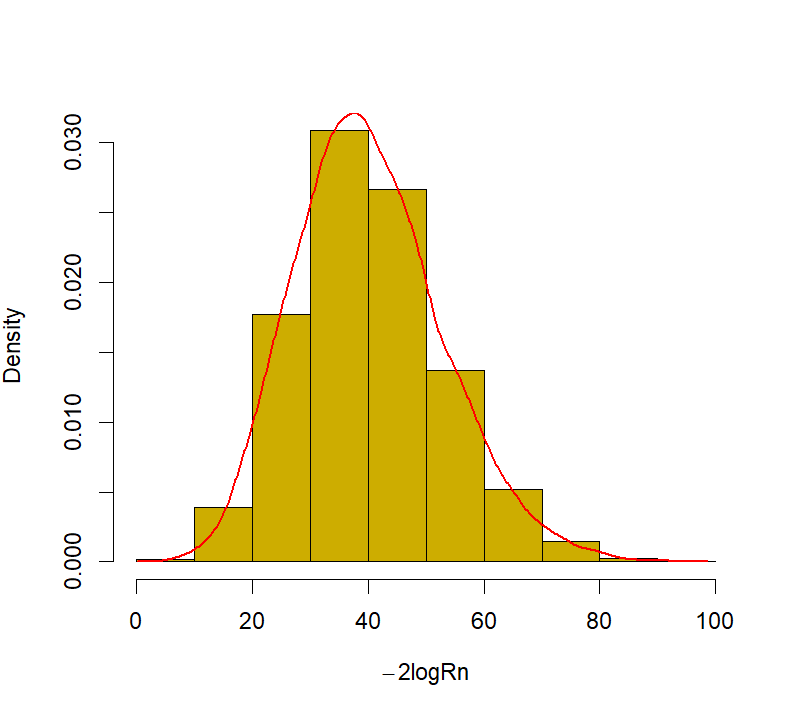} \hskip 0.1cm
\includegraphics[width=8cm,height=6cm]{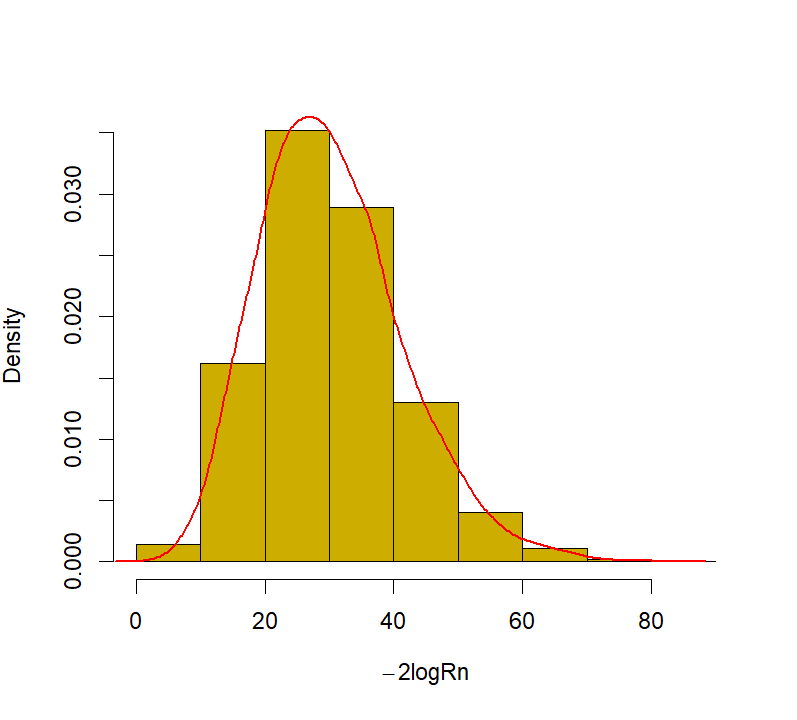}  
\end{center}

$\mathcal{G}_n(\boldsymbol{\tau}=(0.3, 0.2, 0.4), \boldsymbol{\lambda} =(1, 4, 4))$ \hskip 2cm $\mathcal{G}_n(\boldsymbol{\tau}=(0.3, 0.2, 0.4), \boldsymbol{\lambda} =(1, 3, 4))$

\begin{center}
\includegraphics[width=8cm,height=6cm]{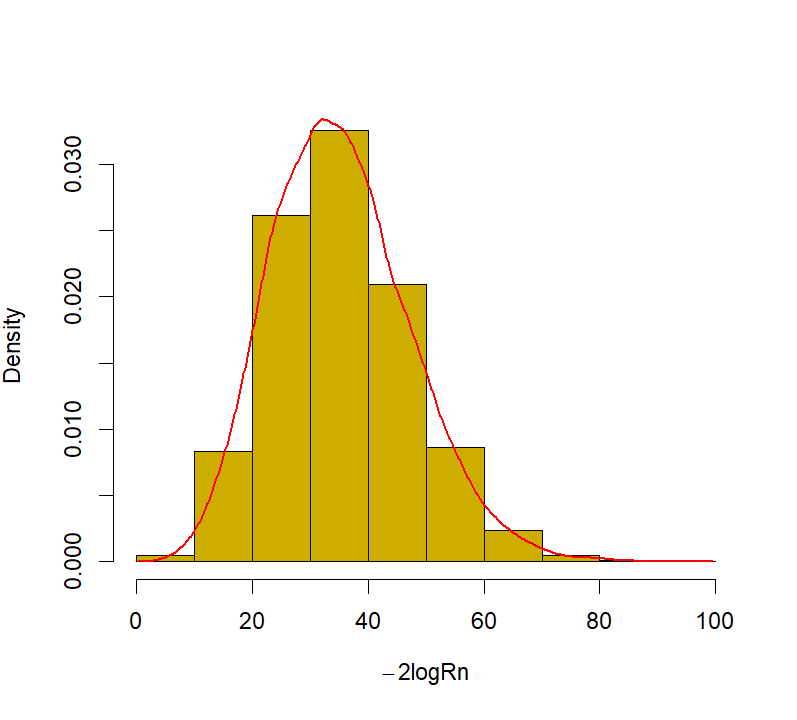} \hskip 0.1cm
\includegraphics[width=8cm,height=6cm]{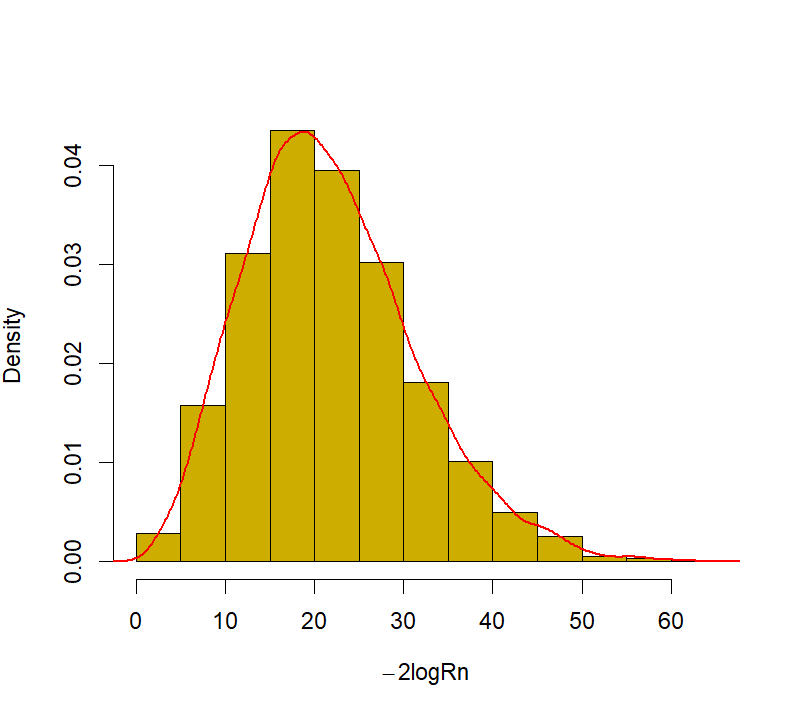}  
\end{center}

$ \mathcal{G}_n(\boldsymbol{\tau}=(0.4, 0.3, 0.2), \boldsymbol{\lambda} =(1, 3, 4))$ \hskip 2cm $\mathcal{G}_n(\boldsymbol{\tau}=(0.2, 0.3, 0.4), \boldsymbol{\lambda} =(1, 3, 4))$

\caption[]{Monte Carlo Approximation Under Alternative Hypothesis}
\label{example2_alt}
\end{figure}

Four scenarios for Example~\ref{example1} are summarized in Table~\ref{example2ThreeLayerPLOTscom}. The first column specifies three configurations of the null hypothesis, while the third column contains four alternative settings. These scenarios are used to examine the empirical distributions of the empirical likelihood (EL) test statistic under both the null and the alternative.
Figure~\ref{example2_null} presents the Monte Carlo approximations under the null hypothesis, and Figure~\ref{example2_alt} shows the corresponding results under the alternative. The conclusions are consistent with those of Example~\ref{example0}. Under the null hypothesis, the empirical densities of the EL test statistic closely follow the $\chi^2_{q=1}$ distribution, providing numerical support for Theorem~\ref{theoremEL}. Under the alternative, the EL test statistic exhibits an approximately normal distribution, in accordance with the Central Limit Theorem.

\begin{table}
\centering

\begin{tabular}{ |c|| c|| c| c |c|}
\hline

 \multicolumn{5}{|c|}{$L = 3$, $r =2$} \\

\hline
\multicolumn{5}{|c|}{ \textbf{Scenario 1}} \\
\hline

$\boldsymbol{\tau}$ & $(0.3, 0.2, 0.4)$ & $(0.3, 0.2, 0.4)$ & $(0.3, 0.2, 0.4)$ & $(0.3, 0.2, 0.4)$ \\
\hline 
$\boldsymbol{\lambda}$ & $(0.8, 0.8, 0.8)$ & $(0.8, 0.7, 0.7)$ & $(0.8, 0.6, 0.6)$ & $(0.8, 0.5, 0.5)$ \\
\hline 


\multicolumn{5}{|c|}{ \textbf{Scenario 2}} \\
\hline
$\boldsymbol{\tau}$ & $(0.3, 0.2, 0.4)$ & $(0.3, 0.2, 0.4)$ & $(0.3, 0.2, 0.4)$ & $(0.3, 0.2, 0.4)$ \\
\hline 
$\boldsymbol{\lambda}$ & $(0.8, 0.8, 0.8)$ & $(0.8, 0.7, 0.6)$ & $(0.8, 0.7, 0.5)$ & $(0.8, 0.6, 0.5)$ \\
\hline


\multicolumn{5}{|c|}{ \textbf{Scenario 3}} \\
\hline
$\boldsymbol{\tau}$ & $(0.4, 0.3, 0.2)$ & $(0.4, 0.3, 0.2)$ & $(0.4, 0.3, 0.2)$ & $(0.4, 0.3, 0.2)$ \\
\hline 
$\boldsymbol{\lambda}$ & $(0.8, 0.8, 0.8)$ & $(0.8, 0.7, 0.6)$ & $(0.8, 0.7, 0.5)$ & $(0.8, 0.6, 0.5)$ \\
\hline 


\multicolumn{5}{|c|}{ \textbf{Scenario 4}} \\
\hline
$\boldsymbol{\tau}$ & $(0.2, 0.3, 0.4)$ & $(0.2, 0.3, 0.4)$ & $(0.2, 0.3, 0.4)$ & $(0.2, 0.3, 0.4)$ \\
\hline 
$\boldsymbol{\lambda}$ & $(0.8, 0.8, 0.8)$ & $(0.8, 0.7, 0.6)$ & $(0.8, 0.7, 0.5)$ & $(0.8, 0.6, 0.5)$ \\
\hline


\end{tabular}

\caption{Four Scenarios of 3-Layer Multilayer Networks by Example \ref{example0}} \label{example1ThreeLayerPLOTscom}
\end{table}

\begin{table}
\centering
\begin{tabular}{ |c|| c|| c| c |c|}
\hline

\multicolumn{5}{|c|}{Three-Layer Networks $L = 3$} \\
\hline

\multicolumn{5}{|c|}{ \textbf{Scenario 1}} \\
\hline
$\boldsymbol{\tau}$ & $(0.3, 0.2, 0.4)$ & $(0.3, 0.2, 0.4)$ & $(0.3, 0.2, 0.4)$ & $(0.3, 0.2, 0.4)$ \\
\hline 
$\boldsymbol{\beta}$ & $(1, 1, 1)$ & $(1, 2, 2)$ & $(1, 3, 3)$ & $(1, 4, 4)$ \\
\hline


\multicolumn{5}{|c|}{ \textbf{Scenario 2}} \\

\hline
$\boldsymbol{\tau}$ & $(0.3, 0.2, 0.4)$ & $(0.3, 0.2, 0.4)$ & $(0.3, 0.2, 0.4)$ & $(0.3, 0.2, 0.4)$ \\
\hline 
$\boldsymbol{\beta}$ & $(1, 1, 1)$ & $(1, 2, 3)$ & $(1, 2, 4)$ & $(1, 3, 4)$ \\
\hline

\multicolumn{5}{|c|}{ \textbf{Scenario 3}} \\

\hline
$\boldsymbol{\tau}$ & $(0.4, 0.3, 0.2)$ & $(0.4, 0.3, 0.2)$ & $(0.4, 0.3, 0.2)$ & $(0.4, 0.3, 0.2)$ \\
\hline 
$\boldsymbol{\beta}$ & $(1, 1, 1)$ & $(1, 2, 3)$ & $(1, 2, 4)$ & $(1, 3, 4)$ \\
\hline


\multicolumn{5}{|c|}{ \textbf{Scenario 4}} \\

\hline
$\boldsymbol{\tau}$ & $(0.2, 0.3, 0.4)$ & $(0.2, 0.3, 0.4)$ & $(0.2, 0.3, 0.4)$ & $(0.2, 0.3, 0.4)$ \\
\hline 
$\boldsymbol{\beta}$ & $(1, 1, 1)$ & $(1, 2, 3)$ & $(1, 2, 4)$ & $(1, 3, 4)$ \\
\hline

\end{tabular}

\caption{Four Scenarios of 3-Layer Multilayer Networks by Example \ref{example1}} \label{example2ThreeLayerPLOTscom}
\end{table}

\section{Simulation Studies}\label{SimRD}

In this section, we study the performance of the proposed EL test on simulated multilayer networks.

\subsection{Simulation}

To evaluate the empirical performance of the EL test, we simulate 1,000 replications under both the null and alternative hypotheses. Under the null hypothesis $H_0: W_1 = \dots = W_L$, the proportion of replications in which $H_0$ is rejected serves as an estimate of the simulated type I error rate, representing the probability of incorrectly rejecting the null. Conversely, under the alternative hypothesis $H_1: W_{l_1} \neq W_{l_2}$ for some $l_1 \neq l_2$, the rejection proportion provides an estimate of the simulated power, which reflects the probability of correctly detecting the presence of differences across layers. 
In this simulation study, we set the nominal (asymptotic) type I error rate to 0.05. We simulate multilayer networks with the number of layers \( L \in \{2, 3, 4\} \) and network sizes \( n \in \{200, 250, 300, 350, 400\} \). For each layer \( l \), the parameter \( \rho_l \) is defined as \( \rho_l = n^{\tau_l} \), where \( \boldsymbol{\tau} = (\tau_1, \dots, \tau_L) \) and \( \tau_l \in \{0.3, 0.2, 0.4, 0.1\} \), ensuring that each \( \tau_l < 0.5 \).


In the first simulation, we consider simulation setting  specified in Example \ref{example0}. 
We take $r\in\{1.5, 2,2.5,3\}$. Denote $\boldsymbol{\tau}=(\tau_1,\dots,\tau_L)$ and $\boldsymbol{\lambda}=(\lambda_1,\dots,\lambda_L)$ with $\lambda_l\in\{0.8,0.7,0.6,0.5\}$. 
Under $H_0$, $\lambda_1=\dots=\lambda_L=0.8$. That is, all the $\lambda_l$ are equal, and hence $W_l$ are the same. Under $H_1$, we consider two cases. The first case corresponds to $\lambda_1=0.8$ and $\lambda_2=\dots=\lambda_L\in\{0.7,0.6,0.5\}$. In the second case, $\lambda_1=0.8$, $\lambda_2,\dots,\lambda_L\in\{0.7,0.6,0.5\}$ and $\lambda_2,\dots,\lambda_L$ are not the equal. 
The simulation results are presented in Table \ref{ELexample1-2Layers} for two layers networks, in Table \ref{ELexample1-3Layers} and Table \ref{ELexample1-3Layers-1} for three layers networks, in Table \ref{ELexample1-4Layers} and Table \ref{ELexample1-4Layers-1} for four layers networks. 
For each multilayer networks indicated by $\boldsymbol{\tau}$ and $\boldsymbol{\lambda}$ in these tables, the corresponding hypothesis is also stated. The $Difference$ associated with each hypothesis is also reported. $Difference$ is calculated as follows 
\begin{equation}\label{differencelambda}
\text{$Difference$} = \sum_{l} |\lambda_1 - \lambda_l|, \quad \text{where } \lambda_l \in \boldsymbol{\lambda}.
\end{equation}

The simulated type I errors are listed in the second columns of Tables \ref{ELexample1-2Layers}--\ref{ELexample1-4Layers-1} under $W_1 = \dots = W_L$. Majority of the results are close to 0.05, indicating that Theorem \ref{theoremEL} works well for small network size $n$. As $n$ get larger, the performance of Type I error gets better. 
For fixed $r, L$ and $\lambda$, the power increases as the network size $n$ increases. For fixed $r, L$ and $n$, the power increases as the $Difference$ gets larger. Moreover, the maximum power is almost one. These findings indicate the consistency of the power of the EL test.
The power values in \textbf{bold} in Tables \ref{ELexample1-2Layers}--\ref{ELexample1-4Layers-1} indicate that the EL test achieves higher or equal power than the WDDT test in our earlier work \cite{yuan-yao2025testing} under the same settings.


In the second simulation, the networks are generated from the model specified in Example \ref{example1}. Let $\boldsymbol{\beta}=(\beta_1,\beta_2,\beta_3,\beta_4)$ with $\beta_l\in\{1,2,3,4\}$.
Under $H_0$,  $\beta_1=\dots=\beta_L=1$. In this case, $W_1=\dots=W_L$. Under $H_1$, there are two scenarios. In the first scenario, $\beta_1=1$ and $\beta_2=\dots=\beta_L\in\{2,3,4\}$. In the second scenario, $\beta_1=1$, $\beta_2,\dots,\beta_L\in\{2,3,4\}$ and  $\beta_2,\dots,\beta_L$ are not equal. 
Table \ref{ELexample2TwoLayer} shows the results of two layer networks, Table \ref{ELexample2ThreeLayer} shows the results of three layer networks, Table \ref{ELexample2FourLayer} shows the results of four layer networks. 
For each multilayer networks indicated by $\boldsymbol{\tau}$ and $\boldsymbol{\beta}$ presented in the tables, the corresponding hypothesis is stated below the networks. 
The $Difference$ associated with each hypothesis is also reported in tables. $Difference$ is calculated as follows 
\begin{equation}\label{differencebeta}
\text{$Difference$} = \sum_{l} |\beta_1 - \beta_l|, \quad \text{where } \beta_l \in \boldsymbol{\beta}.
\end{equation}

The second columns in Tables \ref{ELexample2TwoLayer}--\ref{ELexample2FourLayer} show most of the type I errors are close to 0.05 when $n \geq 300$. This result indicates Theorem \ref{theoremEL} works for small network size $n$ and performs better when $n$ gets larger. 
For fixed $ L,\boldsymbol{\beta}$, the power increases as the network size $n$ increases. For fixed $L,n$, the power increases as the $Difference$ gets larger. Moreover, the maximum power is one. These findings indicate that the power of the EL test is consistent.
The power values in \textbf{bold} in Tables~\ref{ELexample2TwoLayer}--\ref{ELexample2FourLayer} indicate that the EL test achieves higher or equal power than the WDDT test under the same settings. The EL test exhibits obvious advantages over the WDDT test in our earlier work \cite{yuan-yao2025testing} on the terms of powers.

\begin{table}
\centering
\small 
\begin{tabular}{|c || c|| c| c |c|}
\hline
\multicolumn{5}{|c|}{EL test: Two-Layer Networks $L = 2$} \\
\hline
\multicolumn{5}{|c|}{$\boldsymbol{r=1.5}$} \\

\hline
$\boldsymbol{\tau}$ & (0.3, 0.2) & (0.3, 0.2) & (0.3, 0.2) & (0.3, 0.2) \\
\hline
$\boldsymbol{\lambda}$ & (0.8, 0.8) & (0.8, 0.7) & (0.8, 0.6) & (0.8, 0.5) \\
\hline

Hypotheses & 
 $W_1= W_2$
 & $W_1\neq W_2$ & $W_1\neq W_2$ & $W_1\neq W_2$  \\
 \hline
$Difference$ & $0$ & $0.1$ & $0.2$ & $0.3$ \\
\hline
$n=250$ & $0.050$ &$0.070$ &$0.299$ &$0.807$ \\
\hline
$n=300$ & $0.053$ &$0.082$ &$0.377$ &
$0.896$ \\
\hline
\end{tabular}

\caption{EL test for 2-Layer Multilayer Networks by Example \ref{example0}: One Difference in $H_1$ }
\label{ELexample1-2Layers}
\end{table}

\begin{table}
\centering
\small 

\begin{tabular}{|c || c|| c| c |c|}
 \hline
 \multicolumn{5}{|c|}{EL test: Three-Layer Networks $L = 3$} \\
 \hline

\multicolumn{5}{|c|}{$\boldsymbol{r=1.5}$} \\
\hline

$\boldsymbol{\tau}$ & (0.3, 0.2, 0.4) & (0.3, 0.2, 0.4) & (0.3, 0.2, 0.4) & (0.3, 0.2, 0.4) \\
\hline
$\boldsymbol{\lambda}$ & (0.8, 0.8, 0.8) & (0.8, 0.7, 0.7) & (0.8, 0.6, 0.6) & (0.8, 0.5, 0.5) \\
\hline
Hypotheses & 
 $W_1= W_2= W_3$
 & $W_1\neq W_2= W_3$ & $W_1\neq W_2= W_3$ & $W_1\neq W_2= W_3$  \\
\hline

$Difference$ & $0$ & $0.2$ & $0.4$ & $0.6$ \\
\hline
$n=250$& $0.056$ &$0.082$ &$0.564$ &$0.982$ \\
\hline
$n=300$& $0.060$ &$0.110$ &$0.702$ &
$\boldsymbol{0.998}$ \\
\hline

\end{tabular}
\caption{EL test for 3-Layer Multilayer Networks by Example \ref{example0}: One Difference in $H_1$}\label{ELexample1-3Layers}
\end{table}

\begin{table}
\centering
\small

\begin{tabular}{|c || c|| c| c |c|}
\hline
\multicolumn{5}{|c|}{EL test: Three-Layer Networks $L = 3$} \\
\hline
\multicolumn{5}{|c|}{$\boldsymbol{r=1.5}$} \\

\hline
$\boldsymbol{\tau}$ & (0.3, 0.2, 0.4) & (0.3, 0.2, 0.4) & (0.3, 0.2, 0.4) & (0.3, 0.2, 0.4) \\
\hline
$\boldsymbol{\lambda}$ & (0.8, 0.8, 0.8) & (0.8, 0.7, 0.6) & (0.8, 0.7, 0.5) & (0.8, 0.6, 0.5) \\
\hline
Hypotheses & 
 $W_1= W_2= W_3$
 & $W_1\neq W_2\neq W_3$ & $W_1\neq W_2\neq W_3$ & $W_1\neq W_2\neq W_3$  \\
\hline

$Difference$ & $0$ & $0.3$ & $0.4$ & $0.5$ \\
\hline
$n=250$& $0.056$ &$0.274$ &
$\boldsymbol{0.650}$ &
$0.876$ \\
\hline
$n=300$& $0.060$ &$0.342$ &
$\boldsymbol{0.807}$ &
$\boldsymbol{0.957}$ \\
\hline

\end{tabular}

\caption{EL test for 3-Layer Multilayer Networks by Example \ref{example0}: More Than One Difference in $H_1$}\label{ELexample1-3Layers-1}
\end{table}

\begin{table}
\centering
\scriptsize

\begin{tabular}{|c || c|| c| c |c|}
 \hline
 \multicolumn{5}{|c|}{EL test: Four-Layer Networks $L = 4$} \\
 \hline

\multicolumn{5}{|c|}{$\boldsymbol{r=1.5}$} \\
\hline
$\boldsymbol{\tau}$ & (0.3, 0.2, 0.4, 0.1) & (0.3, 0.2, 0.4, 0.1) & (0.3, 0.2, 0.4, 0.1) & (0.3, 0.2, 0.4, 0.1) \\
\hline
$\boldsymbol{\lambda}$ & (0.8, 0.8, 0.8, 0.8) & (0.8, 0.7, 0.7, 0.7) & (0.8, 0.6, 0.6, 0.6) & (0.8, 0.5, 0.5, 0.5) \\
\hline

 Hypotheses & 
 $W_1= W_2= W_3 = W_4$
 & $W_1\neq W_2= W_3= W_4$ & $W_1\neq W_2= W_3= W_4$ & $W_1\neq W_2= W_3= W_4$ \\
\hline

$Difference$ & $0$ & $0.3$ & $0.6$ & $0.9$ \\
\hline
$n=300$& $0.058$ &$0.089$ &$0.609$ &
$\boldsymbol{0.995}$ \\
\hline
$n=350$ & $0.053$ &$0.105$ &$0.739$ &
$\boldsymbol{1.000}$ \\
\hline

\end{tabular}
\caption{EL test for 4-Layer Multilayer Networks by Example \ref{example0}: One Difference in $H_1$}\label{ELexample1-4Layers}
\end{table}

\begin{table}
\centering
\scriptsize

\begin{tabular}{|c || c|| c| c |c|}
\hline
\multicolumn{5}{|c|}{EL test: Four-Layer Networks $L = 4$} \\
\hline

\multicolumn{5}{|c|}{$\boldsymbol{r=1.5}$} \\
\hline
$\boldsymbol{\tau}$ & (0.3, 0.2, 0.4, 0.1) & (0.3, 0.2, 0.4, 0.1) & (0.3, 0.2, 0.4, 0.1) & (0.3, 0.2, 0.4, 0.1) \\
\hline
$\boldsymbol{\lambda}$ & (0.8, 0.8, 0.8, 0.8) & (0.8, 0.7, 0.6, 0.6) & (0.8, 0.7, 0.6, 0.5) & (0.8, 0.7, 0.5, 0.5) \\
 \hline
 
Hypotheses & $W_1= W_2= W_3 = W_4$
 & $W_1 \neq W_2 \neq W_3 = W_4$ & $W_1 \neq W_2 \neq W_3 \neq W_4$ & $W_1 \neq W_2 \neq W_3 = W_4$ \\
\hline

$Difference$ & $0$ & $0.5$ & $0.6$ & $0.7$ \\
\hline
$n=300$& $0.058$ &$0.411$ &$0.696$ &$0.880$ \\
\hline
$n=350$ & $0.053$ &$0.476$ &$0.805$ & $\boldsymbol{0.950}$ \\
\hline
\end{tabular}

\caption{EL test for 4-Layer Multilayer Networks by Example \ref{example0}: More Than One Difference in $H_1$}\label{ELexample1-4Layers-1}
\end{table}

\begin{table}
\centering
\small 
\begin{tabular}{ |c|| c|| c| c |c|}
\hline
 
\multicolumn{5}{|c|}{EL test: Two-Layer Networks $L = 2$} \\
\hline
$\boldsymbol{\tau}$ & (0.3, 0.2) & (0.3, 0.2) & (0.3, 0.2) & (0.3, 0.2) \\
\hline
$\boldsymbol{\beta}$ & (1, 1) & (1, 2) & (1, 3) & (1, 4) \\
 \hline
 
Hypotheses & $W_1= W_2$
 & $W_1 \neq W_2$ & $W_1 \neq W_2$ & $W_1 \neq W_2$ \\
 \hline
 
$Difference$ & $0$ & $1$ & $2$ & $3$ \\
\hline
$n=300$& $0.059$ &$0.173$ & 
$\boldsymbol{0.778}$ & 
$\boldsymbol{0.984}$ \\
\hline
$n=350$ & $0.057$ &$0.213$ & 
$\boldsymbol{0.855}$ & 
$\boldsymbol{0.996}$ \\
\hline
$n=400$ & $0.058$ &$0.258$ & 
$\boldsymbol{0.909}$ & 
$\boldsymbol{1.000}$ \\
\hline
\end{tabular}

\caption{EL test for 2-Layer Multilayer Networks by Example \ref{example1}} \label{ELexample2TwoLayer}
\end{table}

\begin{table}
\centering
\small 
\begin{tabular}{ |c|| c|| c| c |c|}
\hline

\multicolumn{5}{|c|}{EL test: Three-Layer Networks $L = 3$} \\
\hline
\multicolumn{5}{|c|}{ One Difference in $H_1$} \\
\hline
$\boldsymbol{\tau}$ & (0.3, 0.2, 0.4) & (0.3, 0.2, 0.4) & (0.3, 0.2, 0.4) & (0.3, 0.2, 0.4) \\
\hline
$\boldsymbol{\beta}$ & (1, 1, 1) & (1, 2, 2) & (1, 3, 3) & (1, 4, 4) \\
 \hline

 Hypotheses& 
 $W_1= W_2= W_3$
 & $W_1\neq W_2= W_3$ & $W_1\neq W_2= W_3$ & $W_1\neq W_2= W_3$  \\
 \hline
 
$Difference$ & $0$ & $2$ & $4$ & $6$ \\
\hline
$n=300$& $0.063$ &$0.347$ & 
$\boldsymbol{0.989}$ & 
$\boldsymbol{1.000}$ \\
\hline
$n=350$ & $0.058$ &$0.438$ & 
$\boldsymbol{1.000}$ & 
$\boldsymbol{1.000}$ \\
\hline
$n=400$ & $0.059$ &$0.528$ & 
$\boldsymbol{1.000}$ & 
$\boldsymbol{1.000}$ \\
\hline

\multicolumn{5}{|c|}{ More Than One Difference in $H_1$} \\
\hline
$\boldsymbol{\tau}$ & (0.3, 0.2, 0.4) & (0.3, 0.2, 0.4) & (0.3, 0.2, 0.4) & (0.3, 0.2, 0.4) \\
\hline
$\boldsymbol{\beta}$ & (1, 1, 1) & (1, 2, 3) & (1, 2, 4) & (1, 3, 4) \\
 \hline

Hypotheses & 
 $W_1= W_2= W_3$
 & $W_1\neq W_2\neq W_3$ & $W_1\neq W_2\neq W_3$ & $W_1\neq W_2\neq W_3$  \\
 \hline 
 
$Difference$ & $0$ & $3$ & $4$ & $5$ \\
\hline 
$n=300$& $0.063$ &$0.856$ & 
$\boldsymbol{0.996}$ & 
$\boldsymbol{1.000}$ \\
\hline
$n=350$ & $0.058$ &$0.930$ & 
$\boldsymbol{1.000}$ & 
$\boldsymbol{1.000}$ \\
\hline
$n=400$ & $0.059$ &$0.982$ & 
$\boldsymbol{1.000}$ & 
$\boldsymbol{1.000}$ \\
\hline

\end{tabular}
\caption{EL test for 3-Layer Multilayer Networks by Example \ref{example1}} \label{ELexample2ThreeLayer}
\end{table}

\begin{table}
\centering
\scriptsize
\begin{tabular}{ |c|| c|| c| c |c|}
\hline

 \multicolumn{5}{|c|}{EL test: Four-Layer Networks $L = 4$} \\
 \hline
 \multicolumn{5}{|c|}{One Difference in $H_1$} \\
 
\hline
$\boldsymbol{\tau}$ & (0.3, 0.2, 0.4, 0.1) & (0.3, 0.2, 0.4, 0.1) & (0.3, 0.2, 0.4, 0.1) & (0.3, 0.2, 0.4, 0.1) \\
\hline
$\boldsymbol{\beta}$ & (1, 1, 1, 1) & (1, 2, 2, 2) & (1, 3, 3, 3) & (1, 4, 4, 4) \\
 \hline

Hypotheses & 
 $W_1= W_2= W_3 = W_4$
 & $W_1\neq W_2= W_3= W_4$ & $W_1\neq W_2= W_3= W_4$ & $W_1\neq W_2= W_3= W_4$ \\
\hline

$Difference$ & $0$ & $3$ & $6$ & $9$ \\
\hline
$n=300$& $0.070$ &$0.297$ & 
$\boldsymbol{0.954}$ & 
$\boldsymbol{1.000}$ \\
\hline
$n=350$ & $0.065$ &$0.370$ &
$\boldsymbol{0.990}$ & 
$\boldsymbol{1.000}$ \\
\hline
$n=400$ & $0.058$ &$0.458$ & 
$\boldsymbol{0.998}$& 
$\boldsymbol{1.000}$ \\
\hline

\multicolumn{5}{|c|}{More Than One Difference in $H_1$} \\
\hline
$\boldsymbol{\tau}$ & (0.3, 0.2, 0.4, 0.1) & (0.3, 0.2, 0.4, 0.1) & (0.3, 0.2, 0.4, 0.1) & (0.3, 0.2, 0.4, 0.1) \\
\hline
$\boldsymbol{\beta}$ & (1, 1, 1, 1) & (1, 2, 3, 3) & (1, 2, 3, 4) & (1, 2, 4, 4) \\
\hline

Hypotheses & $W_1= W_2= W_3 = W_4$
 & $W_1 \neq W_2 \neq W_3= W_4$ & $W_1 \neq W_2 \neq W_3 \neq W_4$ & $W_1 \neq W_2 \neq W_3 = W_4$ \\
\hline

$Difference$ & $0$ & $5$ & $6$ & $7$ \\
\hline
$n=300$& $0.070$ &$0.830$ & 
$\boldsymbol{0.949}$ & 
$\boldsymbol{0.988}$ \\
\hline
$n=350$ & $0.065$ &$0.912$ &$0.982$ & $\boldsymbol{0.998}$ \\
\hline
$n=400$ & $0.058$ &$0.958$ & 
$\boldsymbol{0.997}$ & 
$\boldsymbol{1.000}$ \\
\hline

\end{tabular}
\caption{EL test for 4-Layer Multilayer Networks by Example \ref{example1}} \label{ELexample2FourLayer}
\end{table}

\subsection{Factors that Impact Performance of EL Test}\label{ELpermute}

Each scenario in Table~\ref{example1ThreeLayerPLOTscom} for Example~\ref{example0} and in Table~\ref{example2ThreeLayerPLOTscom} for Example~\ref{example1} consists of four multilayer networks. To assess how the configurations of the multilayer networks affect the testing performance, we cyclically permute the order of layers within each multilayer network in each scenario, treating each network in turn as the first layer, and then evaluate the resulting test outcomes. Starting from the original ordering $(A_1, A_2, A_3)$, we consider the following cyclic permutations: $(A_2, A_3, A_1)$ and $(A_3, A_1, A_2)$. We define one permutation set as
\begin{equation}\label{permutation}
\left\{ (A_1, A_2, A_3), \; (A_2, A_3, A_1), \; (A_3, A_1, A_2) \right\}.
\end{equation}


Firstly, permutations of each multilayer networks under each scenario in Table~\ref{example1ThreeLayerPLOTscom} for Example \ref{example0} are tested. The corresponding results by EL test are reported in Tables \ref{ELexample1L3-P1}--\ref{ELexample1L3-P4} for each scenario. The second column in these tables are the four multilayer networks in each scenario, the third and fourth columns contain cyclic permutation. The highest power values within each permutation are shown in \textbf{bold}. 
We observe that multilayer networks with larger values of $Difference$, $\tau_1$, and $\lambda_1$ in the first layer tend to yield higher simulated power. When multilayer networks share the same $Difference$ value, a larger $\tau_1$ in the first layer appears to be more critical than a larger $\lambda_1$ for achieving higher power. Furthermore, even when the $Difference$ value is smaller, the presence of the largest $\tau_1$ value in the first layer still results in the highest observed power. These findings suggest that the $\tau_1$ value in the first layer plays a more influential role than the $Difference$ value and $\lambda_1$ value in determining the power of the test. 
By observing Table~\ref{ELexample1L3-P2}, the configuration $(\tau_1 = 0.4, \lambda_1 = 0.6)$ yields higher power than $(\tau_1 = 0.3, \lambda_1 = 0.8)$ given the same $Difference$. In Table~\ref{ELexample1L3-P4}, $(\tau_1 = 0.4, \lambda_1 = 0.5)$ yields larger power than $(\tau_1 = 0.2, \lambda_1 = 0.8)$ even though $(\tau_1 = 0.4, \lambda_1 = 0.5)$ is associated with a smaller $Difference$. It can be inferred that a larger $\tau_1$ in the first layer has a strong impact in guaranteeing higher powers for multilayer networks.

Accordingly, we list six networks---characterized by their $\tau$ and $\lambda$ parameters---that can serve as the first layer in multilayer networks in Table~\ref{SixNetworks_Visual1}. These networks are listed by their potential simulated powers. 
Multilayer networks with the first layer \((\tau_1 = 0.4, \lambda_1 = 0.8)\) achieve the highest power, followed by \((\tau_1 = 0.4, \lambda_1 = 0.6)\) and \((\tau_1 = 0.4, \lambda_1 = 0.5)\). Since it is observed that larger $\tau_1$ is more important than other factors, \((\tau_1 = 0.4, \lambda_1 = 0.5)\) is followed by \((\tau_1 = 0.3, \lambda_1 = 0.8)\), \((\tau_1 = 0.2, \lambda_1 = 0.8)\). Multilayer networks with the first layer \((\tau_1 = 0.1, \lambda_1 = 0.5)\) yield the lowest power. 

\begin{figure}
\begin{center}
\includegraphics[width=14cm,height=8cm]{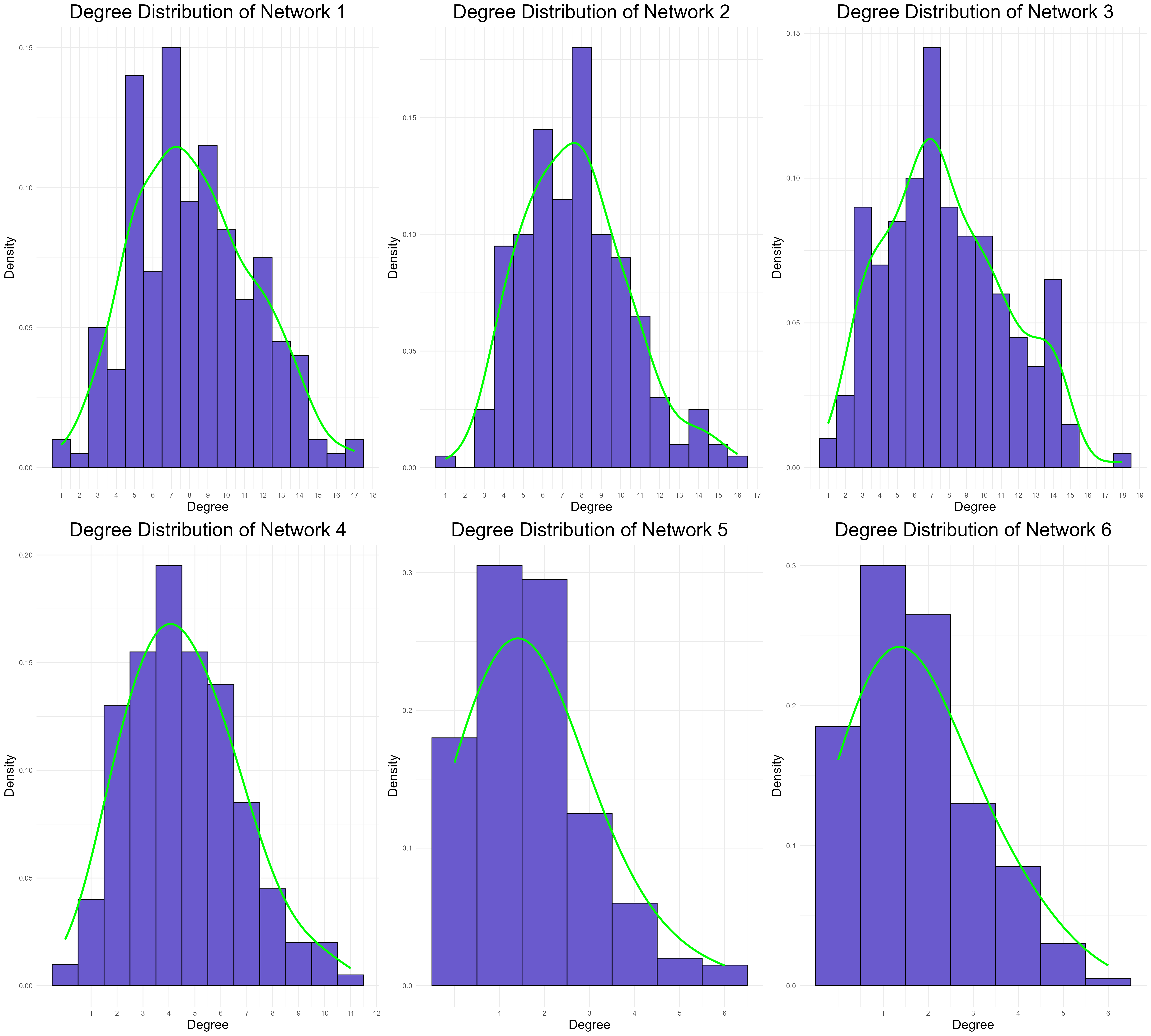}
\end{center}
\caption[Degree Distributions of Networks  $\sim \mathcal{G}_n(\boldsymbol{\tau}, \boldsymbol{\lambda})$]{Degree Distributions of Networks $ \sim \mathcal{G}_n(\boldsymbol{\tau}=(0.4, 0.4, 0.4, 0.3, 0.2, 0.1), \boldsymbol{\lambda} =(0.8, 0.6, 0.5, 0.8, 0.8, 0.5))$}
\label{example1_D_SIX}
\end{figure}

By examining the properties of these networks, we can identify which types of multilayer networks tend to exhibit higher simulated power under the EL test. Moreover, this understanding may inform predictions about whether the EL test will demonstrate strong power when applied to real data with similar structural characteristics. 
To support this analysis, we first visualize the network structures and then compute key network metrics, including density, total degree, average degree, degree distribution, clustering coefficient, number of connected components, and path length. 
The results for the networks from Example~\ref{example0} are presented in Table~\ref{SixNetworks_Visual1}. Degree distributions are visualized in Figures~\ref{example1_D_SIX}. 
We observe that large $\tau$ and $\lambda$ values generate moderately dense networks, and multilayer networks with a high-density first layer tend to achieve higher simulated power. The degree distributions are approximately bell-shaped, with increasing right skewness observed as the network sparsity increases.

\begin{table}
\centering

\begin{tabular}{|c || c| c| c | c| c| c|}
\hline
\multicolumn{7}{|c|}{Six Networks $\boldsymbol{r=2, n=200}$} \\
\hline
$\boldsymbol{\tau}$ & $0.4$ & $0.4$ & $0.4$ & $0.3$ & $0.2$ & $0.1$ \\
\hline 
$\boldsymbol{\lambda}$ & $0.8$ & $0.6$ & $0.5$ & $0.8$ & $0.8$ & $0.5$ \\
\hline 

Density & $0.0417$ &$0.0416$ &$0.0388$ &$0.0244$ &$0.0148$ &$0.0075$   \\
\hline
Total Degree & $1660$ &$1654$ &$1546$ &$972$ &$588$ &$298$   \\
\hline
Average Degree & $8.30$ &$8.27$ &$7.73$ &$4.86$ &$2.94$ &$1.49$   \\
\hline
Average Clustering Coefficient & $0.0369$ &$0.0464$ &$0.0413$ &$0.0275$ &$0.0146$ &$0.0000$   \\
\hline
Connected Components & $1$ &$1$ &$1$ &$3$ &$8$ &$62$   \\
\hline
Path Length (Diameter) & $4$ &$5$ &$5$ &$7$ &$11$ &$18$   \\
\hline

\end{tabular}
\caption{Characteristics of Simulated Networks by Example \ref{example0}}\label{SixNetworks_Visual1}
\end{table}

For Example \ref{example1}, permutations of each multilayer network under each scenario in Table~\ref{example2ThreeLayerPLOTscom} are tested. The corresponding results are reported in Tables \ref{ELexample2L3-P1}--\ref{ELexample2L3-P4} for each scenario. The highest power values within each permutation are shown in \textbf{bold}. 
We observe the following. In Scenario 1, multilayer networks across all permutations with the highest $Difference$ values tend to yield the highest power. In Scenario 2, when multilayer networks have the same $Difference$ value, the configuration with $(\tau_1 = 0.3, \beta_1 = 1)$ exhibits higher power than the one with $(\tau_1 = 0.4, \beta_1 = 3)$. However, the network with $(\tau_1 = 0.3, \beta_1 = 0.8)$ by  Example~\ref{example0} has lower power than the one with $(\tau_1 = 0.4, \beta_1 = 0.6)$, suggesting that the $\tau$ value plays a more critical role in determining power in Example~\ref{example0} than in Example~\ref{example1}. In Scenario 3, the multilayer network with $(\tau_1 = 0.4, \beta_1 = 1)$ achieves the highest power even when it has a relatively smaller $Difference$ value. In Scenario 4, when $Difference$ values are equal, the configuration $(\tau_1 = 0.4, \beta_1 = 3)$ yields higher power than $(\tau_1 = 0.2, \beta_1 = 1)$. Comparing with Example~\ref{example0}, the configuration $(\tau_1 = 0.4, \lambda_1 = 0.5)$ can retain higher power, whereas $(\tau_1 = 0.4, \beta_1 = 4)$ in Example~\ref{example1} does not. This indicates that a larger $\tau$ value in the first layer is more important for achieving higher power in Example~\ref{example0} than in Example~\ref{example1}. 
In summary, we find that multilayer networks with a larger $\tau$ value and a smaller $\beta$ value in the first layer, combined with a larger $Difference$, tend to produce higher simulated power. Among these factors, a larger $Difference$ appears to play a more critical role than a larger $\tau$ in the first layer for achieving high power. We observe that the multilayer network with first layer parameters $(\tau_1 = 0.4, \beta_1 = 3)$ yields lower power than the configuration with $(\tau_1 = 0.3, \beta_1 = 1)$ in Table~\ref{ELexample2L3-P2}, but higher power than the configuration with $(\tau_1 = 0.2, \beta_1 = 1)$ in Table~\ref{ELexample2L3-P4}. This comparison reinforces the importance of the interaction between parameters of $\tau$, $\beta$.

Accordingly, we list six networks—based on their $\tau$ and $\beta$ parameters—that can serve as the first layer in multilayer networks in Table~\ref{SixNetworks_Visual2}. 
Multilayer networks with the first layer $(\tau_1 = 0.4, \beta_1 = 1)$ exhibit the highest power and followed by $(\tau_1 = 0.3, \beta_1 = 1)$, $(\tau_1 = 0.4, \beta_1 = 3)$, $(\tau_1 = 0.2, \beta_1 = 1)$. Multilayer networks with the first layer $(\tau_1 = 0.1, \beta_1 = 4)$ yield the lowest.

\begin{figure}
\begin{center}
\includegraphics[width=14cm,height=8cm]{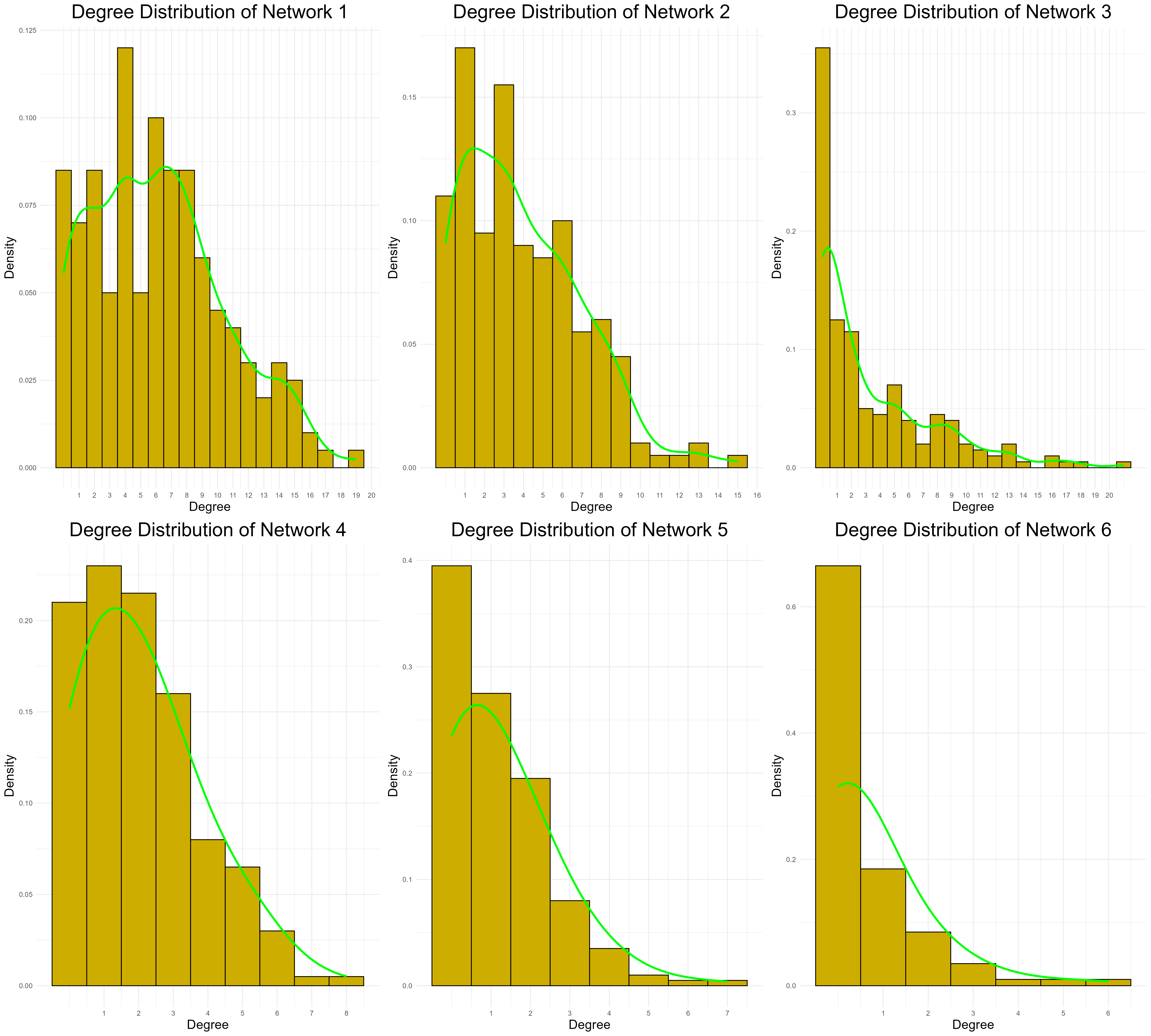}
\end{center}
\caption[Degree Distributions of Networks  $\sim \mathcal{G}_n(\boldsymbol{\tau}, \boldsymbol{\beta})$]
{Degree Distributions of Networks $\sim \mathcal{G}_n(\boldsymbol{\tau}=(0.4, 0.3, 0.4, 0.2, 0.1, 0.1), \boldsymbol{\beta} =(1, 1, 3, 1, 1, 4))$}
\label{example2_D_SIX}
\end{figure}

The results of key network metrics, including density, total degree, average degree, degree distribution, clustering coefficient, number of connected components, and path length from Example~\ref{example1} are shown in Table~\ref{SixNetworks_Visual2}. Degree distributions are visualized in Figures~\ref{example2_D_SIX}. 
Networks generated with large $\tau$ values tend to be moderately dense. Simulation results indicate that multilayer networks with a dense first layer consistently achieve higher power. The degree distributions are right-skewed, with only a small portion of the left tail present, and they become increasingly right-skewed as the network becomes sparser.

\begin{table}
\centering

\begin{tabular}{|c || c| c| c | c| c| c|}
\hline
\multicolumn{7}{|c|}{Six Networks $\boldsymbol{n=200}$} \\
\hline
$\boldsymbol{\tau}$ & $0.4$ & $0.3$ & $0.4$ & $0.2$ & $0.1$ & $0.1$ \\
\hline 
$\boldsymbol{\beta}$ & $1$ & $1$ & $3$ & $1$ & $1$ & $4$ \\
\hline

Density & $0.0311$ &$0.0196$ &$0.0170$ &$0.0103$ &$0.0058$ &$0.0031$   \\
\hline
Total Degree & $1236$ &$780$ &$678$ &$408$ &$232$ &$122$   \\
\hline
Average Degree & $6.18$ &$3.9$ &$3.39$ &$2.04$ &$1.16$ &$0.61$   \\
\hline
Average Clustering Coefficient & $0.0501$ &$0.0336$ &$0.0816$ &$0.0238$ &$0.0324$ &$0.0300$   \\
\hline
Connected Components & $18$ &$23$ &$73$ &$45$ &$92$ &$143$   \\
\hline
Path Length (Diameter) & $6$ &$8$ &$7$ &$12$ &$19$ &$12$   \\
\hline

\end{tabular}
\caption{Characteristics of Simulated Networks by Example \ref{example1}}\label{SixNetworks_Visual2}
\end{table}

\begin{table}
\centering
\begin{tabular}{ |c|| c| c| c |}
 \hline
\multicolumn{4}{|c|}{EL test: Three-Layer Networks $L =3, r=2$} \\
\hline
$\boldsymbol{\tau}$ & $(0.3, 0.2, 0.4)$ & $(0.2, 0.4, 0.3)$ & $(0.4, 0.3, 0.2)$ \\
\hline 
$\boldsymbol{\lambda}$ & $(0.8, 0.8, 0.8)$ & $(0.8, 0.8, 0.8)$ & $(0.8, 0.8, 0.8)$ \\
\hline

Hypotheses & $W_1= W_2= W_3$ & $W_2= W_3= W_1$ & $W_3= W_1= W_2$ \\
\hline
 
$Difference$ & $0$ & $0$ & $0$ \\
 \hline
$n=200$& $0.060$ &$0.055$ &$0.065$  \\
\hline
$n=250$& $0.056$ &$0.059$ &$0.064$  \\
\hline
$n=300$& $0.044$ &$0.058$ &$0.057$  \\
\hline
\end{tabular}
\caption{EL test for 3-Layer Multilayer Networks by Example \ref{example0}: Scenario 1} \label{ELexample1L3-P1}
\end{table}


\begin{table}
\centering

\begin{tabular}{ |c|| c| c| c |}
\hline
\multicolumn{4}{|c|}{EL test: Three-Layer Networks $L =3, r=2$} \\
\hline
$\boldsymbol{\tau}$ & $(0.3, 0.2, 0.4)$ & $(0.2, 0.4, 0.3)$ & $(0.4, 0.3, 0.2)$ \\
\hline 
$\boldsymbol{\lambda}$ & $(0.8, 0.8, 0.8)$ & $(0.8, 0.8, 0.8)$ & $(0.8, 0.8, 0.8)$ \\
\hline

Hypotheses & $W_1= W_2= W_3$ & $W_2= W_3= W_1$ & $W_3= W_1= W_2$ \\
\hline
 
$Difference$ & $0$ & $0$ & $0$ \\
\hline
$n=200$ & $0.060$ & $0.055$ & $0.065$  \\
\hline
$n=250$ & $0.056$ & $0.059$ & $0.064$  \\
\hline
$n=300$ & $0.044$ & $0.058$ & $0.066$  \\
\hline
\end{tabular}
\caption{EL test for 3-Layer Multilayer Networks by Example \ref{example0}: Scenario 2} \label{ELexample1L3-P2}
\end{table}

\begin{table}
\centering


\begin{tabular}{ |c|| c| c| c |}
\hline
\multicolumn{4}{|c|}{EL test: Three-Layer Networks $L =3, r=2$} \\
\hline
$\boldsymbol{\tau}$ & $(0.4, 0.3, 0.2)$ & $(0.3, 0.2, 0.4)$ & $(0.2, 0.4, 0.3)$ \\
\hline 
$\boldsymbol{\lambda}$ & $(0.8, 0.8, 0.8)$ & $(0.8, 0.8, 0.8)$ & $(0.8, 0.8, 0.8)$ \\
\hline

Hypotheses & $W_1= W_2= W_3$ & $W_2= W_3= W_1$ & $W_3= W_1= W_2$ \\
\hline

$Difference$ & $0$ & $0$ & $0$ \\
\hline
$n=200$ & $0.065$ & $0.060$ & $0.055$  \\
\hline
$n=250$ & $0.064$ & $0.056$ & $0.059$  \\
\hline
$n=300$ & $0.057$ & $0.044$ & $0.058$  \\
\hline
\end{tabular}
\caption{EL test for 3-Layer Multilayer Networks by Example \ref{example0}: Scenario 3} \label{ELexample1L3-P3}
\end{table}


\begin{table}
\centering


\begin{tabular}{ |c|| c| c| c |}
\hline
 \multicolumn{4}{|c|}{EL test: Three-Layer Networks $L =3, r=2$} \\
\hline
$\boldsymbol{\tau}$ & $(0.2, 0.3, 0.4)$ & $(0.3, 0.4, 0.2)$ & $(0.4, 0.2, 0.3)$ \\
\hline 
$\boldsymbol{\lambda}$ & $(0.8, 0.8, 0.8)$ & $(0.8, 0.8, 0.8)$ & $(0.8, 0.8, 0.8)$ \\
\hline

Hypotheses & $W_1= W_2= W_3$ & $W_2= W_3= W_1$ & $W_3= W_1= W_2$ \\
\hline
 
 $Difference$ & $0$ & $0$ & $0$ \\
 \hline
$n=200$ & $0.056$ & $0.061$ & $0.074$  \\
\hline
$n=250$ & $0.051$ & $0.056$ & $0.066$  \\
\hline
$n=300$ & $0.054$ & $0.058$ & $0.061$  \\
\hline
\end{tabular}
\caption{EL test for 3-Layer Multilayer Networks by Example \ref{example0}: Scenario 4} \label{ELexample1L3-P4}
\end{table}


\begin{table}
\centering
\begin{tabular}{ |c|| c| c| c |}
 \hline
\multicolumn{4}{|c|}{EL test: Three-Layer Networks $L =3$} \\
\hline
$\boldsymbol{\tau}$ & $(0.3, 0.2, 0.4)$ & $(0.2, 0.4, 0.3)$ & $(0.4, 0.3, 0.2)$  \\
\hline 
$\boldsymbol{\beta}$ & $(1, 1, 1)$ & $(1, 1, 1)$ & $(1, 1, 1)$ \\
\hline

Hypotheses & $W_1= W_2= W_3$ & $W_2= W_3= W_1$ & $W_3= W_1= W_2$ \\
\hline

$Difference$ & $0$ & $0$ & $0$ \\
 \hline
$n=300$& $0.063$ &$0.060$ &$0.069$  \\
\hline
$n=350$& $0.058$ &$0.059$ &$0.068$  \\
\hline
$n=400$& $0.059$ &$0.055$ &$0.059$  \\
\hline
\end{tabular}
\caption{EL test for 3-Layer Multilayer Networks by Example \ref{example1}: Scenario 1} \label{ELexample2L3-P1}
\end{table}

\begin{table}
\centering


\begin{tabular}{ |c|| c| c| c |}
\hline
 \multicolumn{4}{|c|}{EL test: Three-Layer Networks $L =3, r=2$} \\
\hline
$\boldsymbol{\tau}$ & $(0.3, 0.2, 0.4)$ & $(0.2, 0.4, 0.3)$ & $(0.4, 0.3, 0.2)$  \\
\hline 
$\boldsymbol{\beta}$ & $(1, 1, 1)$ & $(1, 1, 1)$ & $(1, 1, 1)$ \\
\hline

Hypotheses & $W_1= W_2= W_3$ & $W_2= W_3= W_1$ & $W_3= W_1= W_2$ \\
\hline
 
$Difference$ & $0$ & $0$ & $0$ \\
\hline
$n=300$& $0.063$ &$0.060$ &$0.069$  \\
\hline
$n=350$& $0.058$ &$0.059$ &$0.068$  \\
\hline
$n=400$& $0.059$ &$0.055$ &$0.059$  \\
\hline
\end{tabular}
\caption{EL test for 3-Layer Multilayer Networks by Example \ref{example1}: Scenario 2} \label{ELexample2L3-P2}
\end{table}


\begin{table}
\centering


\begin{tabular}{ |c|| c| c| c |}
\hline
 \multicolumn{4}{|c|}{EL test: Three-Layer Networks $L =3, r=2$} \\
\hline
$\boldsymbol{\tau}$ & $(0.4, 0.3, 0.2)$ & $(0.3, 0.2, 0.4)$ & $(0.2, 0.4, 0.3)$  \\
\hline 
$\boldsymbol{\beta}$ & $(1, 1, 1)$ & $(1, 1, 1)$ & $(1, 1, 1)$ \\
\hline

Hypotheses & $W_1= W_2= W_3$ & $W_2= W_3= W_1$ & $W_3= W_1= W_2$ \\
\hline
 
$Difference$ & $0$ & $0$ & $0$ \\
\hline
$n=300$ & $0.069$ & $0.063$ & $0.060$  \\
\hline 
$n=350$ & $0.068$ & $0.058$ & $0.059$  \\
\hline
$n=400$ & $0.059$ & $0.059$ & $0.055$  \\
\hline
\end{tabular}
\caption{EL test for 3-Layer Multilayer Networks by Example \ref{example1}: Scenario 3} \label{ELexample2L3-P3}
\end{table}


\begin{table}
\centering

\begin{tabular}{ |c|| c| c| c |}
\hline
\multicolumn{4}{|c|}{EL test: Three-Layer Networks $L =3$} \\
\hline
$\boldsymbol{\tau}$ & $(0.2, 0.3, 0.4)$ & $(0.3, 0.4, 0.2)$ & $(0.4, 0.2, 0.3)$  \\
\hline 
$\boldsymbol{\beta}$ & $(1, 1, 1)$ & $(1, 1, 1)$ & $(1, 1, 1)$ \\
\hline

Hypotheses & $W_1= W_2= W_3$ & $W_2= W_3= W_1$ & $W_3= W_1= W_2$ \\
\hline
 
$Difference$ & $0$ & $0$ & $0$ \\
\hline
$n=300$ & $0.061$ & $0.069$ & $0.070$ \\
\hline 
$n=350$ & $0.059$ & $0.068$ & $0.074$ \\
\hline
$n=400$ & $0.055$ & $0.063$ & $0.060$ \\
\hline
\end{tabular}
\caption{EL test for 3-Layer Multilayer Networks by Example \ref{example1}: Scenario 4} \label{ELexample2L3-P4}
\end{table}

\subsection{Robustness of EL test under Rank-2 Multilayer Networks}\label{ELrank2simulations}

To evaluate the robustness of the proposed WDDT test, we conduct simulations for networks with rank-2 multilayer networks. A rank-2 matrix has at most two \emph{linearly independent} columns \cite{rencher1997matrix, axler2024linear}. We first define two linear independent vectors by \textit{degree-correction parameters} $W_l$ in \textit{Rank-1 Random Multilayer Heterogeneous Graphs} model: \[W_{l1} = [w_1, w_2, \dots, w_{\frac{n}{2}}, 0, \dots, 0]^\top,\] and \[W_{l2} = [0, 0, \dots, 0, w_{\frac{n}{2}+1}, \dots, w_{n}]^\top.\]
Then, generate rank-2 multilayer networks using the following linear algebra in (\ref{rank2})
\begin{equation}\label{rank2}
\mathbb{E}[A_l] =
\begin{bmatrix}
W_{l1}, W_{l2}
\end{bmatrix} 
\begin{bmatrix}
a \rho_l & b \rho_l \\
b \rho_l & a \rho_l \\
\end{bmatrix} 
\begin{bmatrix}
W_{l1}^\top \\
W_{l2}^\top \\
\end{bmatrix}.
\end{equation}
After expansion, it is \[ \mathbb{E}[A_l] = a\rho_l W_{l1} W_{l1}^\top + b\rho_l W_{l2} W_{l1}^\top  + b \rho_l W_{l1} W_{l2}^\top + a \rho_l W_{l2} W_{l2}^\top,\] from which we can see the number of \textit{linear independent} columns in adjacency matrix A is 2, which are $W_{l1}$ and $W_{l2}$. Therefore, rank of adjacency matrix A is 2. 
More specifically, \textit{Rank-2 Random Multilayer Heterogeneous Graphs} for simulation can equivalently be written as
\begin{equation}\label{rank2eq}
\mathbb{P}(A_{l;ij}=1) =
\begin{cases}
1.1\rho_l W_{l,i} W_{l,j}, & \text{if } 1 \leq i,j \leq \frac{n}{2}, \ \text{or } \frac{n}{2} < i,j \leq n, \\
0.9\rho_l W_{l,i} W_{l,j}, & \text{if } 1 \leq i \leq \frac{n}{2} < j \leq n,
\end{cases}
\end{equation}
where $a=1.1$, $b=0.9$, $\rho_l$ and $W_{l,i}$ are specified in Example~\ref{example0} and Example~\ref{example1}.


Firstly, we consider Example \ref{example0}. Four scenarios for three-layer networks in Table~\ref{example1ThreeLayerPLOTscom} are evaluated for rank-2 multilayer networks. The permutations of each multilayer network in these four scenarios are evaluated, and the results are reported in Tables \ref{ELexample1L3-P1-Rank2}--\ref{ELexample1L3-P4-Rank2}. The highest power values within each permutation are shown in \textbf{bold}. 
Majority of the type I errors are close to 0.05. As $n$ increases, the power gets larger and approaches 1. All results demonstrate that the EL test is robust for rank-2 multilayer networks. Same with rank-1 multilayer networks, rank-2 multilayer networks with larger $\tau_1$ and $\lambda_1$ values, along with a greater \textit{Difference} value, tend to produce higher statistical power. Accordingly, the six networks listed in Table~\ref{SixNetworks_Visual1} remain valid in the rank-2 setting.


Secondly, four scenarios for three-layer networks of Exmaple \ref{example1} in Table~\ref{example2ThreeLayerPLOTscom} are evaluated for rank-2 multilayer networks. The permutations of each multilayer network in these four scenarios are evaluated, and the results are reported in Tables \ref{ELexample2L3-P1-Rank2}--\ref{ELexample2L3-P4-Rank2}. The highest power values within each permutation are shown in \textbf{bold}. The type I errors are close to 0.05. As $n$ increases, the power gets larger and approaches 1. All results demonstrate that the EL test is robust for rank-2 multilayer networks.
As observed in rank-1 multilayer networks, rank-2 multilayer networks with larger values of $\tau_1$ and $\beta_1$, as well as a greater \textit{Difference} value, tend to exhibit higher power. Accordingly, the six networks listed in Table~\ref{SixNetworks_Visual2} are also valid in the rank-2 setting.


\begin{table}
\centering


\begin{tabular}{ |c|| c| c| c |}
 \hline
\multicolumn{4}{|c|}{EL test: Three-Layer Networks $L =3, r=2$} \\
\hline
$\boldsymbol{\tau}$ & $(0.3, 0.2, 0.4)$ & $(0.2, 0.4, 0.3)$ & $(0.4, 0.3, 0.2)$ \\
\hline 
$\boldsymbol{\lambda}$ & $(0.8, 0.8, 0.8)$ & $(0.8, 0.8, 0.8)$ & $(0.8, 0.8, 0.8)$ \\
\hline

Hypotheses & $W_1= W_2= W_3$ & $W_2= W_3= W_1$ & $W_3= W_1= W_2$ \\
\hline

$Difference$ & $0$ & $0$ & $0$ \\
 \hline
$n=250$& $0.059$ &$0.063$ &$0.071$  \\
\hline
$n=300$& $0.061$ &$0.060$ &$0.066$  \\
\hline
$n=350$& $0.045$ &$0.059$ &$0.052$  \\
\hline
\end{tabular}
\caption{EL test for 3-Layer Multilayer Networks of Rank-2 by Example \ref{example0}: Scenario 1} \label{ELexample1L3-P1-Rank2}
\end{table}

\begin{table}
\centering


\begin{tabular}{ |c|| c| c| c |}
\hline
 \multicolumn{4}{|c|}{EL test: Three-Layer Networks $L =3, r=2$} \\
\hline
$\boldsymbol{\tau}$ & $(0.3, 0.2, 0.4)$ & $(0.2, 0.4, 0.3)$ & $(0.4, 0.3, 0.2)$ \\
\hline 
$\boldsymbol{\lambda}$ & $(0.8, 0.8, 0.8)$ & $(0.8, 0.8, 0.8)$ & $(0.8, 0.8, 0.8)$ \\
\hline

Hypotheses & $W_1= W_2= W_3$ & $W_2= W_3= W_1$ & $W_3= W_1= W_2$ \\
\hline
 
$Difference$ & $0$ & $0$ & $0$ \\
\hline
$n=250$ & $0.059$ & $0.063$ & $0.071$  \\
\hline
$n=300$ & $0.061$ & $0.060$ & $0.066$  \\
\hline 
$n=350$ & $0.045$ & $0.059$ & $0.052$  \\
\hline
\end{tabular}
\caption{EL test for 3-Layer Multilayer Networks of Rank-2 by Example \ref{example0}: Scenario 2} \label{ELexample1L3-P2-Rank2}
\end{table}

\begin{table}
\centering


\begin{tabular}{ |c|| c| c| c |}
\hline
\multicolumn{4}{|c|}{EL test: Three-Layer Networks $L =3, r=2$} \\
\hline
$\boldsymbol{\tau}$ & $(0.4, 0.3, 0.2)$ & $(0.3, 0.2, 0.4)$ & $(0.2, 0.4, 0.3)$ \\
\hline 
$\boldsymbol{\lambda}$ & $(0.8, 0.8, 0.8)$ & $(0.8, 0.8, 0.8)$ & $(0.8, 0.8, 0.8)$ \\
\hline

Hypotheses & $W_1= W_2= W_3$ & $W_2= W_3= W_1$ & $W_3= W_1= W_2$ \\
\hline

$Difference$ & $0$ & $0$ & $0$ \\
\hline
$n=250$ & $0.071$ & $0.059$ & $0.063$  \\
\hline
$n=300$ & $0.066$ & $0.061$ & $0.060$  \\
\hline 
$n=350$ & $0.052$ & $0.045$ & $0.059$  \\
\hline
\end{tabular}
\caption{EL test for 3-Layer Multilayer Networks of Rank-2 by Example \ref{example0}: Scenario 3} \label{ELexample1L3-P3-Rank2}
\end{table}


\begin{table}
\centering


\begin{tabular}{ |c|| c| c| c |}
\hline
 \multicolumn{4}{|c|}{EL test: Three-Layer Networks $L =3, r=2$} \\
\hline

$\boldsymbol{\tau}$ & $(0.2, 0.3, 0.4)$ & $(0.3, 0.4, 0.2)$ & $(0.4, 0.2, 0.3)$ \\
\hline 
$\boldsymbol{\lambda}$ & $(0.8, 0.8, 0.8)$ & $(0.8, 0.8, 0.8)$ & $(0.8, 0.8, 0.8)$ \\
\hline

Hypotheses & $W_1= W_2= W_3$ & $W_2= W_3= W_1$ & $W_3= W_1= W_2$ \\
\hline
$Difference$ & $0$ & $0$ & $0$ \\
\hline
$n=250$ & $0.058$ & $0.062$ & $0.069$  \\
\hline
$n=300$ & $0.056$ & $0.061$ & $0.062$  \\
\hline 
$n=350$ & $0.055$ & $0.064$ & $0.062$  \\
\hline
\end{tabular}
\caption{EL test for 3-Layer Multilayer Networks of Rank-2 by Example \ref{example0}: Scenario 4} \label{ELexample1L3-P4-Rank2}
\end{table}


\begin{table}
\centering



\begin{tabular}{ |c|| c| c| c |}
 \hline
\multicolumn{4}{|c|}{EL test: Three-Layer Networks $L =3$} \\
\hline

$\boldsymbol{\tau}$ & $(0.3, 0.2, 0.4)$ & $(0.2, 0.4, 0.3)$ & $(0.4, 0.3, 0.2)$  \\
\hline 
$\boldsymbol{\beta}$ & $(1, 1, 1)$ & $(1, 1, 1)$ & $(1, 1, 1)$ \\
\hline

Hypotheses & $W_1= W_2= W_3$ & $W_2= W_3= W_1$ & $W_3= W_1= W_2$ \\
\hline
 
$Difference$ & $0$ & $0$ & $0$ \\
 \hline
$n=300$& $0.063$ &$0.061$ &$0.075$  \\
\hline
$n=350$& $0.063$ &$0.064$ &$0.057$  \\
\hline
$n=400$& $0.059$ &$0.058$ &$0.069$  \\
\hline
\end{tabular}
\caption{EL test for 3-Layer Multilayer Networks of Rank-2 by Example \ref{example1}: Scenario 1} \label{ELexample2L3-P1-Rank2}
\end{table}


\begin{table}
\centering


\begin{tabular}{ |c|| c| c| c |}
\hline
 \multicolumn{4}{|c|}{EL test: Three-Layer Networks $L =3, r=2$} \\
\hline

$\boldsymbol{\tau}$ & $(0.3, 0.2, 0.4)$ & $(0.2, 0.4, 0.3)$ & $(0.4, 0.3, 0.2)$  \\
\hline 
$\boldsymbol{\beta}$ & $(1, 1, 1)$ & $(1, 1, 1)$ & $(1, 1, 1)$ \\
\hline

Hypotheses & $W_1= W_2= W_3$ & $W_2= W_3= W_1$ & $W_3= W_1= W_2$ \\
\hline
 
 $Difference$ & $0$ & $0$ & $0$ \\
 \hline
$n=300$ & $0.063$ & $0.061$ & $0.075$  \\
\hline 
$n=350$ & $0.063$ & $0.064$ & $0.057$  \\
\hline
$n=400$ & $0.059$ & $0.058$ & $0.069$  \\
\hline
\end{tabular}
\caption{EL test for 3-Layer Multilayer Networks of Rank-2 by Example \ref{example1}: Scenario 2} \label{ELexample2L3-P2-Rank2}
\end{table}


\begin{table}
\centering


\begin{tabular}{ |c|| c| c| c |}
\hline
 \multicolumn{4}{|c|}{EL test: Three-Layer Networks $L =3, r=2$} \\
\hline

$\boldsymbol{\tau}$ & $(0.4, 0.3, 0.2)$ & $(0.3, 0.2, 0.4)$ & $(0.2, 0.4, 0.3)$  \\
\hline 
$\boldsymbol{\beta}$ & $(1, 1, 1)$ & $(1, 1, 1)$ & $(1, 1, 1)$ \\
\hline

Hypotheses & $W_1= W_2= W_3$ & $W_2= W_3= W_1$ & $W_3= W_1= W_2$ \\
 \hline
 
 $Difference$ & $0$ & $0$ & $0$ \\
 \hline
$n=300$ & $0.075$ & $0.063$ & $0.064$  \\
\hline 
$n=350$ & $0.057$ & $0.063$ & $0.064$  \\
\hline
$n=400$ & $0.069$ & $0.059$ & $0.058$  \\
\hline
\end{tabular}
\caption{EL test for 3-Layer Multilayer Networks of Rank-2 by Example \ref{example1}: Scenario 3} \label{ELexample2L3-P3-Rank2}
\end{table}


\begin{table}
\centering


\begin{tabular}{ |c|| c| c| c |}
\hline
\multicolumn{4}{|c|}{EL test: Three-Layer Networks $L =3$} \\
\hline

$\boldsymbol{\tau}$ & $(0.2, 0.3, 0.4)$ & $(0.3, 0.4, 0.2)$ & $(0.4, 0.2, 0.3)$  \\
\hline 
$\boldsymbol{\beta}$ & $(1, 1, 1)$ & $(1, 1, 1)$ & $(1, 1, 1)$ \\
\hline

Hypotheses & $W_1= W_2= W_3$ & $W_2= W_3= W_1$ & $W_3= W_1= W_2$ \\
 \hline
 
 $Difference$ & $0$ & $0$ & $0$ \\
 \hline
$n=300$ & $0.061$ & $0.069$ & $0.084$ \\
\hline 
$n=350$ & $0.065$ & $0.062$ & $0.073$ \\
\hline
$n=400$ & $0.060$ & $0.047$ & $0.066$ \\
\hline
\end{tabular}
\caption{EL test for 3-Layer Multilayer Networks of Rank-2 by Example \ref{example1}: Scenario 4} \label{ELexample2L3-P4-Rank2}
\end{table}

\section{Real Data Application}\label{realdata}

In this subsection, we apply the proposed EL test to the multilayer social network CS-Aarhus, available in \cite{MMR13}. The CS-Aarhus networks are undirected, unweighted, and consist of five types of online and offline relationships among the 61 employees of the Computer Science Department at Aarhus University. The five network layers are defined as follows:
\begin{itemize}
    \item $A_1$: Lunch — two individuals have lunch together.
    \item $A_2$: Facebook — two individuals are connected on Facebook.
    \item $A_3$: Co-authorship — individuals have co-authored publications.
    \item $A_4$: Leisure — individuals spend leisure time together.
    \item $A_5$: Work — individuals work together.
\end{itemize}
The five layers $A_1, A_2, A_3, A_4, A_5$ are listed in Table~\ref{WDDTredata1} and visualized in Figure~\ref{cs-realdata1}.

\medskip 

\begin{figure}
\begin{center}
\includegraphics[width=18cm,height=8cm]{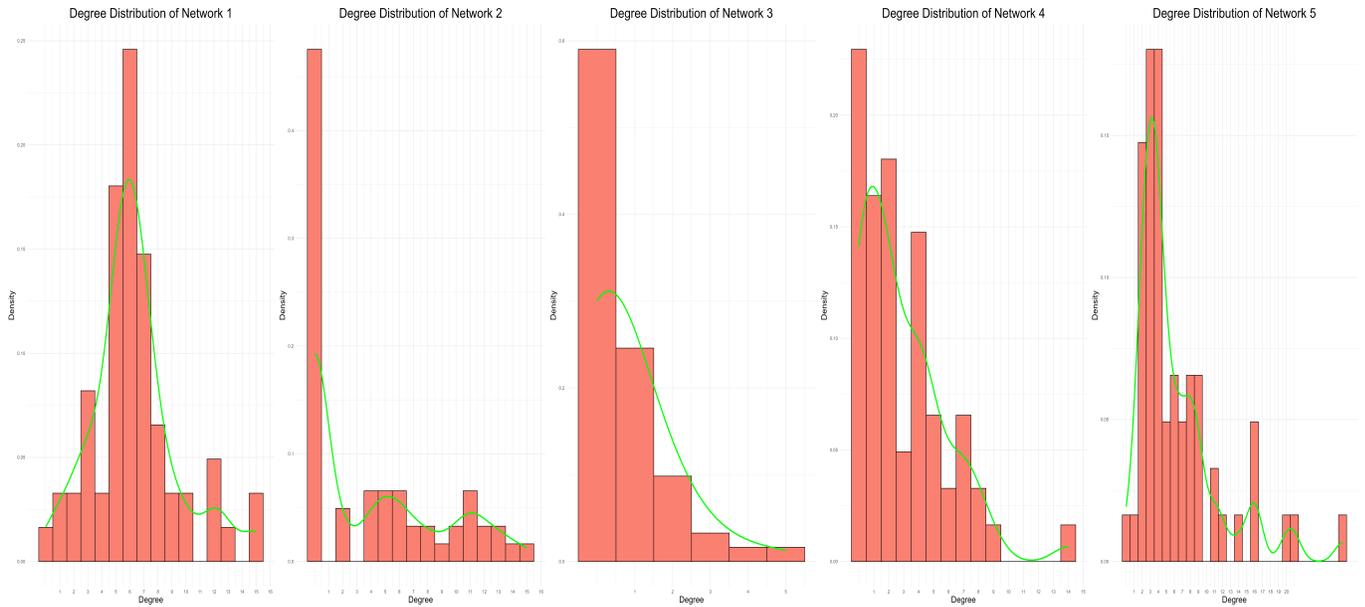}
\end{center}
\caption{Degree Distributions of 5 Layers Real-World Networks}
\label{RealData1_DegreeDistribution}
\end{figure}

Table~\ref{RealData1_property} presents the characteristics of the five-layer real-world network. The characteristics include network density, total degree, average degree, clustering coefficient, number of connected components, and average path length. Based on the density values, we observe that network $A_3$ is the sparsest, while networks $A_1$ and $A_5$ are relatively denser.
Figure~\ref{RealData1_DegreeDistribution} displays the degree distributions of each network layer. The distributions for networks $A_1$ and $A_5$ exhibit bell-shaped curves, which are symmetric and unimodal, centered around the mean degree. This shape indicates that most nodes in the network have degrees close to the average, while very few nodes have either extremely low or extremely high degrees. Such a pattern is characteristic of networks where connectivity is fairly uniform across nodes, as opposed to scale-free networks, which exhibit heavy-tailed or power-law distributions with many low-degree nodes and a few highly connected hubs. In contrast, the degree distributions of networks $A_2$, $A_3$, and $A_4$ resemble the right tail of a bell-shaped curve, with only a small portion of the left tail present. This pattern suggests that the majority of nodes have moderate degrees, while a smaller proportion possess higher connectivity, resulting in a gradually declining right tail. The truncation of the left tail may indicate either a scarcity of very low-degree nodes or an analytical focus on the network's core structure.

\begin{table}
\centering
\begin{tabular}{ |c || c|}
\hline
\multicolumn{2}{|c|}{5 Layers Real-World Networks} \\
\hline

Layers& Association  \\
\hline
$A_1$ &  Representative of two individuals having Lunch together  \\
\hline
$A_2$ &  Representative of two individuals having a social connection via Facebook   \\
\hline
$A_3$ &  Representative of two individuals co-authoring a publication   \\
\hline
$A_4$ &  Representative of two individuals having Leisure together   \\
\hline
$A_5$ &  Representative of two individuals working together   \\
\hline

\end{tabular}

\caption{5 Layers Real-World Network} \label{WDDTredata1}
\end{table}

\begin{table}
\centering

\begin{tabular}{|c || c| c| c | c| c|}
 \hline
 \multicolumn{6}{|c|}{5 Layers Real-World Networks} \\
 \hline

$ $ & $A_1$ & $A_2$ & $A_3$ & $A_4$ & $A_5$ \\ 
\hline

Density & $0.1055$ &$0.0678$ &$0.0115$ &$0.0481$ &$0.1060$   \\
\hline
Total Degree & $386$ &$248$ &$42$ &$176$ &$388$   \\
\hline
Average Degree & $6.328$ &$4.066$ &$0.689$ &$2.885$ &$6.361$   \\
\hline
Clustering Coefficient & $0.5689$ &$0.4806$ &$0.4286$ &$0.3431$ &$0.3388$   \\
\hline
Connected Components & $2$ &$30$ &$44$ &$16$ &$2$   \\
\hline
Path Length(Diameter) & $7$ &$4$ &$3$ &$8$ &$4$   \\
\hline

\end{tabular}
\caption{Characteristics of 5 Layers Real-World Networks}\label{RealData1_property}
\end{table}

The proposed EL test is applied to the multilayer social networks CS-Aarhus with the results summarized in Table~\ref{redataEL1} and Table~\ref{redataEL2}. 
Firstly, we test whether the five networks share the same degree-correction parameters or a one-dimensional common subspace. We calculate the EL test statistic $R_n$ with each network treated as the first network, along with the corresponding $p$-values. The results are shown in Table~\ref{redataEL1}. The $p$-values of both the EL test are all smaller than 0.05, indicating that the five networks do not share the same degree-correction parameters nor a one-dimensional common subspace.
Next, we test whether each quadripartite subset of the networks shares the same degree-correction parameters. According to Table~\ref{redataEL2}, the $p$-values of both the EL test are also smaller than 0.05, suggesting that no quadripartite subset can be embedded into the same one-dimensional common subspace.
Furthermore, we test whether each triple of networks shares the same degree-correction parameters. The $p$-values of the EL test are again all smaller than 0.05, indicating that these triples cannot be embedded into the same one-dimensional common subspace. 
Lastly, we test whether each pair of networks shares the same degree-correction parameters. The $p$-values of the EL test are all smaller than 0.05, indicating that none of the pairs can be embedded into the same one-dimensional common subspace.

\begin{table}
\centering
\small
\begin{tabular}{|c||ccc |}
\hline
Layers & \shortstack{ Test Statistic} & P-Value & Conclusion 
 \\
\hline
$A_1, A_2, A_3, A_4, A_5$ & $27.329$ & $0.0000$ & $Reject H_0$  \\
\hline
$A_2, A_3, A_4, A_5, A_1$ & $41.937$ & $0.0000$ & $Reject H_0$  \\
\hline
$A_3, A_4, A_5, A_1, A_2$ & $10.981$ & $0.0009$ & $Reject H_0$  \\
\hline
$A_4, A_5, A_1, A_2, A_3$ & $28.041$ & $0.0000$ & $Reject H_0$  \\
\hline
$A_5, A_1, A_2, A_3, A_4$ & $40.808$ & $0.0000$ & $Reject H_0$  
 \\
\hline
\end{tabular}
\caption{EL test for 5 layers real-world networks } \label{redataEL1}
\end{table}

\begin{table}
\centering
\begin{tabular}{|c||ccc |}
\hline
Layers & \shortstack{Test Statistic} & P-Value & Conclusion \\
\hline
$A_1, A_2$ & $27.330$ & $0.0000$ & $Reject H_0$ \\
\hline
 $A_1, A_3$ & $6.924$ & $0.0085$ & $Reject H_0$ \\
\hline
$A_1, A_4$ & $5.560$ & $0.0184$ & $Reject H_0$ \\
\hline
 $A_1, A_5$ & $12.806$ & $0.0003$ & $Reject H_0$ \\
\hline
$A_2, A_3$ & $7.332$ & $0.0068$ & $Reject H_0$ \\
\hline
 $A_2, A_4$ & $23.665$ & $0.0000$ & $Reject H_0$ \\
\hline
 $A_2, A_5$ & $32.119$ & $0.0000$ & $Reject H_0$ \\
\hline
 $A_3, A_4$ & $4.629$ & $0.0314$ & $Reject H_0$ \\
\hline
 $A_3, A_5$ & $10.189$ & $0.0014$ & $Reject H_0$ \\
\hline
 $A_4, A_5$ & $29.849$ & $0.0000$ & $Reject H_0$ \\
\hline
\hline 
 $A_3, A_4, A_5$ & $10.164$ & $0.0014$ & $Reject H_0$  \\
\hline
 $A_2, A_4, A_5$ & $46.302$ & $0.0000$ & $Reject H_0$ \\
\hline
 $A_2, A_3, A_5$ & $24.577$ & $0.0000$ & $Reject H_0$  \\
\hline
 $A_2, A_3, A_4$ & $17.833$ & $0.0000$ & $Reject H_0$ \\
\hline
 $A_1, A_4, A_5$ & $17.864$ & $0.0000$ & $Reject H_0$  \\
\hline
 $A_1, A_3, A_5$ & $12.199$ & $0.0005$ & $Reject H_0$  \\
\hline
 $A_1, A_3, A_4$ & $10.335$ & $0.0013$ & $Reject H_0$  \\
\hline
 $A_1, A_2, A_5$ & $30.686$ & $0.0000$ & $Reject H_0$  \\
\hline
 $A_1, A_2, A_4$ & $25.320$ & $0.0000$ & $Reject H_0$  \\
\hline
 $A_1, A_2, A_3$ & $19.538$ & $0.0000$ & $Reject H_0$  \\
\hline
\hline 
 $A_1, A_2, A_3, A_4$ & $21.088$ & $0.0000$ & $Reject H_0$   \\
\hline
 $A_1, A_2, A_3, A_5$ & $26.534$ & $0.0000$ & $Reject H_0$  \\
\hline
 $A_1, A_2, A_4, A_5$ & $36.097$ & $0.0000$ & $Reject H_0$  \\
\hline
 $A_1, A_3, A_4, A_5$ & $15.607$ & $0.0001$ & $Reject H_0$  \\
\hline
 $A_2, A_3, A_4, A_5$ & $36.181$ & $0.0000$ & $Reject H_0$ \\
\hline
\end{tabular}
\caption{EL test for real-world networks } \label{redataEL2}
\end{table}


\section{Discussion}\label{discuss}

Multilayer networks provide a richer and more realistic representation of real-world complex systems than traditional single-layer network. In many real-world scenarios, interactions between entities are multifaceted, and multilayer networks offer a powerful framework for capturing such complexity. As a result, multilayer networks has been widely applied and actively studied. Given a multilayer network, a natural and important question arises: \textit{Does a common subspace exist across all networks?} Answering this question could help in understanding more information extracted across all layers that captures their homogeneity or shared common structure, which have many practical applications. 

In this work, we propose the empirical likelihood ratio (EL) test to assess whether all networks share a common invariant subspace. Under the null hypothesis, all network layers are assumed to share the same subspace, whereas under the alternative hypothesis, only some layers share a common subspace. We conduct comprehensive simulation studies to investigate the limiting distribution and evaluate the performance of the EL test. Monte Carlo approximations confirm the validity of the test, and the simulation results indicate that it performs well and achieves higher power than the weighted degree difference test (WDDT) from our earlier work \cite{yuan-yao2025testing}, which was the first test developed for detecting a common invariant subspace in multilayer networks. These results highlight the advantages of the EL test. Additionally, we apply the EL test to real-world multilayer network data, illustrating its robustness and practical utility.

As a future research proposal on testing common subspace in multilayer networks, we consider more complex random multilayer heterogeneous graphs model of rank-$q$, where $q \geq 2$. \emph{Rank-1 random multilayer heterogeneous graphs} model in Definition \ref{defmg} consider $W_l$ vectors as homogeneity vectors and $\rho_l$ as heterogeneity score. Similarly, we define \emph{rank-q random multilayer heterogeneous graphs} model by define homogeneity matrix $U_{n \times q}$ and heterogeneity matrix $\Lambda_{q \times q}$. The expected adjacency matrices in multilayer networks are decomposed as 
\[\mathbb{E}[A_l] = U_l \Lambda_l U_l^\top .\]

Each $U_l$ is latent invariant subspace of multilayer networks and its property of invariance implies that a linear transformation on $U_l$ is also within this subspace. Thus, we consider an orthogonal matrix $Q_l$ in each layer to accommodate the isomorphic variance of the subspace $U_l$. Given multilayer networks $A_1,A_2,\dots, A_L$, we are interested in testing the following hypotheses

       
\begin{equation}\label{hypoeq-rank2}
\begin{aligned}
H_0:\; & \forall \, l \in \{2,3,\dots,L\}, \ \exists \text{ an orthogonal matrix } Q_l 
         \text{ such that } U_1 = Q_l U_l, \\[6pt]
H_1:\; & \exists \, l_1 \neq l_2 \text{ such that } 
         U_{l_1} \neq Q U_{l_2} \quad \text{for all orthogonal matrices } Q.
\end{aligned}
\end{equation}

Under $H_0$,  the graphs $A_1,A_2,\dots, A_L$ have the same common invariant subspace. Under $H_1$, there exist at least two graphs such that their common invariant subspace are  different. Correspondingly, test statistics under this model should be constructed, and their asymptotic distributions should be derived.

The simulation results for the Empirical Likelihood (EL) test have demonstrated that the EL test performs well for the hypotheses in~(\ref{hypoeq}). However, a rigorous theoretical justification is needed. Therefore, another important topic for future work is to derive the asymptotic distribution of the EL test. Furthermore, it is of interest to extend the EL framework to the hypotheses in~(\ref{hypoeq-rank2}) and to validate its performance through both simulation studies and asymptotic analysis.



\bibliographystyle{plain}
\bibliography{bibTex}

\end{document}